\documentclass[twocolumn,english,pra,floatfix,tightenlines,superscriptaddress,nofootinbib]{revtex4-1}

\usepackage[T1]{fontenc}
\usepackage[latin9]{inputenc}
\usepackage{amsbsy}
\usepackage{amsmath}
\usepackage{amsthm}
\usepackage{amssymb}
\usepackage{graphicx}
\usepackage{comment}
\usepackage[usenames]{color}
\usepackage{hyperref}
\usepackage{subfigure}
\usepackage{times}
\hypersetup{
    colorlinks=true,       
    linkcolor=cyan,          
    citecolor=magenta,        
    filecolor=magenta,      
    urlcolor=cyan,           
    runcolor=cyan
}

\makeatletter

\newcommand{\hc}{\mathrm{h.c.}}

\newcommand{\al}{\alpha}
\newcommand{\1}{\leavevmode{\rm 1\ifmmode\mkern  -4.8mu\else\kern -.3em\fi I}}
\newcommand{\qt}[1]{\textquotedblleft#1\textquotedblright}

\newcommand{\ignore}[1]{}

\newcommand{\beq}{\begin{equation}}
\newcommand{\eeq}{\end{equation}}
\newcommand{\beqnn}{\begin{equation*}}
\newcommand{\eeqnn}{\end{equation*}}
\newcommand{\bea}{\begin{eqnarray}}
\newcommand{\eea}{\end{eqnarray}}
\newcommand{\beann}{\begin{eqnarray*}}
\newcommand{\eeann}{\end{eqnarray*}}
\newcommand{\bes} {\begin{subequations}}
\newcommand{\ees} {\end{subequations}}
\newcommand{\ketbra}[1]{|{#1}\rangle\langle#1|}
\newcommand{\ket}[1]{|#1\rangle}
\newcommand{\bra}[1]{\langle#1|}
\newcommand{\mK}{\mathcal{K}}
\newcommand{\mD}{\mathcal{D}}
\newcommand{\mL}{\mathcal{L}}

\newtheorem{claim}{Claim}

\makeatother

\usepackage{babel}
\begin{document}

\title{Relaxation \textit{vs}. adiabatic quantum 
steady 
state preparation: which wins?}

\author{Lorenzo Campos Venuti}

\affiliation{Department of Physics \& Astronomy, University of Southern California, Los Angeles,
CA 90089-0484, USA}
\affiliation{Center for Quantum Information
Science \& Technology, University of Southern California, Los Angeles,
CA 90089-0484, USA}

\author{Tameem Albash}

\affiliation{Department of Physics \& Astronomy, University of Southern California, Los Angeles,
CA 90089-0484, USA}
\affiliation{Center for Quantum Information
Science \& Technology, University of Southern California, Los Angeles,
CA 90089-0484, USA}
\affiliation{Information Sciences Institute, University of Southern California,
Marina del Rey, California 90292, USA}

\author{Milad Marvian}

\affiliation{Center for Quantum Information
Science \& Technology, University of Southern California, Los Angeles,
CA 90089-0484, USA}
\affiliation{Department of Electrical Engineering, University of Southern California, Los Angeles,
CA 90089-0484, USA}

\author{Daniel Lidar}

\affiliation{Department of Physics \& Astronomy, University of Southern California, Los Angeles,
CA 90089-0484, USA}
\affiliation{Center for Quantum Information
Science \& Technology, University of Southern California, Los Angeles,
CA 90089-0484, USA}
\affiliation{Department of Electrical Engineering, University of Southern California, Los Angeles,
CA 90089-0484, USA}
\affiliation{Department of Chemistry, University of Southern California, Los Angeles,
CA 90089, USA}

\author{Paolo Zanardi}

\affiliation{Department of Physics \& Astronomy, University of Southern California, Los Angeles,
CA 90089-0484, USA}
\affiliation{Center for Quantum Information
Science \& Technology, University of Southern California, Los Angeles,
CA 90089-0484, USA}
\begin{abstract}
Adiabatic preparation of the ground states of many-body Hamiltonians in the closed system limit is at the heart of adiabatic quantum computation, but in reality systems are always open. This motivates a natural comparison between, on the one hand, adiabatic preparation of steady states of Lindbladian generators and, on the other hand, relaxation towards the same steady states subject to the final Lindbladian of the adiabatic process. 
In this work we thus adopt the perspective that the goal is the most efficient possible preparation of such steady states, rather than ground states.
Using known rigorous bounds for the open-system adiabatic theorem and for mixing times, we are 
then
led to a disturbing conclusion that at first appears to doom efforts to build physical quantum annealers: relaxation seems to always converge {faster} than
adiabatic preparation. However, by carefully estimating the adiabatic preparation time  for Lindbladians describing thermalization in the low temperature limit, we show that there is, after all, room for an adiabatic speedup over relaxation. To test the analytically derived bounds for the adiabatic preparation time and the relaxation time, we numerically study three models: a dissipative quasi-free fermionic chain, a single qubit coupled to a thermal bath, and the \qt{spike} problem of $n$ qubits coupled to a thermal bath. Via these models we find that the answer to the \qt{which wins} question depends for each model on the temperature and the system-bath coupling strength. In the case of the \qt{spike} problem we find that relaxation during the adiabatic evolution plays an important role in ensuring a speedup over the final-time relaxation procedure. Thus, relaxation-assisted adiabatic preparation can be more efficient than both pure adiabatic evolution and pure relaxation.

\end{abstract}


\maketitle

\section{Introduction}

Encoding the result of a computation into the ground state of a real or simulated physical system is a powerful idea that is shared by quantum approaches such as quantum annealing (QA) \cite{kadowaki_quantum_1998} and adiabatic quantum computing (AQC) \cite{farhi_quantum_2000}, and classical relaxation heuristics based on Monte Carlo Markov chains (MCMC), such as simulated annealing \cite{kirkpatrick_optimization_1983} and parallel tempering \cite{Hukushima:1996}. Unfortunately, comparisons between these quantum and classical approaches, while necessary and worthwhile, are fraught with difficulties at the outset, because they are formulated in very different settings. While MCMC algorithms are implemented on digital classical computers, adiabatic preparation (via QA or AQC) is an inherently analog process implemented on quantum hardware. Moreover, the choice of the specific classical heuristic, the CPU architecture, and so on, all contribute with unknown scaling factors and polynomial overhead, turning a fair comparison into a subtle and challenging task. Additionally, in all but a few cases the best possible algorithms are unknown, leading to the need for a careful distinction between different types of quantum speedups \cite{ronnow_defining_2014}. 

Instead of attempting to compare quantum and classical approaches, one may ask whether the quantum analog of classical relaxation heuristics, namely quantum relaxation, is a viable alternative to adiabatic preparation. Here we provide a systematic study of this question and its answer. We shall argue that the comparison between adiabatic preparation and quantum relaxation is natural and allows us to put the two approaches on an equal footing. 

In order to carry out this program, we assume that the system dynamics can be described by a time-dependent Liouvillian generator of Lindblad type ${\cal L}(t)$ \cite{Lindblad:76}, with $t\in[0,\tau]$. Adiabatic preparation and quantum relaxation are described in a unified framework as follows. The target for both strategies is the  steady state $\rho_{\mathrm{SS}} (\tau)$ of $\mathcal{L}(\tau)$, i.e., the state that satisfies $\mathcal{L}(\tau)[\rho_{\mathrm{SS}}(\tau)]=0$. 
It is important to clarify that by focusing on steady state preparation we will not address the problem of ground state preparation (the usual goal of optimization via AQC and QA).

Adiabatic preparation is the process of adiabatically evolving an
initial state $\rho_{\mathrm{SS}}(0)$, subject to the generator
$\mathcal{L}(t)$. The adiabatic theorem for open quantum systems
guarantees that the final state $\rho_{\mathrm{adia}}(\tau) =
\mathrm{Texp}\left[\int_{0}^{\tau}\mathcal{L}(t)dt\right][\rho(0)]$
approaches the steady state state $\rho_{\mathrm{SS}} (\tau)$ in the
limit of infinitely slow change of  $\mathcal{L}(t)$
\cite{joye_general_2007,salem_quasi-static_2007,oreshkov_adiabatic_2010,avron_adiabatic_2012,venuti_adiabaticity_2016}. In
contrast, quantum relaxation is described by the dynamics 
$\rho_{\mathrm{relax}}(t) =e^{t\mathcal{L}(\tau)}[\rho(0)]$ where
$\rho(0)$ is the initial state. Results estimating the convergence
rate of 
$\rho_{\mathrm{relax}}(t)$ towards
$\rho_{\mathrm{SS}} (\tau)$ are known
\cite{temme_2-divergence_2010,kastoryano_rapid_2013,kastoryano_quantum_2013}. The
adiabatic preparation of the state $\rho_{\mathrm{SS}} (\tau)$ can now
be naturally compared to the process of quantum relaxation towards the
same state. 

Our approach has a number of potential applications: (i) 
It allows us to study the efficiency of realistic implementations of AQC and QA for which the interaction with the environment cannot be neglected \cite{childs_robustness_2001,sarandy_adiabatic_2005-1,Aberg:2005rt,PhysRevA.74.052330,PhysRevA.75.062313,amin_decoherence_2009,Qiang:13,Albash:2015nx,Wild:2016il}. In such situations we may expect that the open-system dynamics converge (possibly after an exponentially long time)
not to the ground state but rather
to a thermal equilibrium Gibbs state $\rho_{\mathrm{eq}} \sim \exp{(-\beta H)}$, where $\beta=1/T>0$ is the inverse temperature of the system and $H$ its Hamiltonian. (ii) One may also consider idealized dynamics designed to prepare ground states instead of thermal states. In this case $\beta$ becomes an external tunable parameter that one tries to make as large as possible, as in simulated annealing or quantum Monte Carlo algorithms \cite{Werner:2005fb}.
However, there is still a difference in the sense that the thermal state has equal weight on all ground states in the case of degeneracy, whereas closed system adiabatic evolution need not. 
(iii) Davies generators \cite{davies_markovian_1974}, which are a subclass of physically realistic generators, have thermal Gibbs states as steady states \cite{Breuer:book}, and the relaxation process may well be called thermalization in this case. 
If the final Hamiltonian is classical (e.g., diagonal in the
computational basis), the relaxation process is described by a 
classical Markov chain, the so called Pauli equation.\footnote{This
  requires, as we do, that the initial state is diagonal in the Hamiltonian
  eigenbasis. The infinite temperature initial state 
belongs to
this class.} 
Hence, in this situation we end up comparing 
adiabatic quantum preparation with classical Markov algorithms.  
Note, however, that we do not measure the efficiency via the time
necessary to run the algorithm on a digital classical computer as is
usually done, but rather run the classical Markov chain 
on an analog device as well.

For both the adiabatic and the  relaxation approaches the efficiency is
encoded in the \qt{time-to-steady-state} (TTSS). The TTSS is the
minimum time $\tau$ required to be $\epsilon$-close (in an appropriate
distance measure) to the desired steady state.  The time $\tau$ can be
estimated using known results for the adiabatic theorem
\cite{joye_general_2007,salem_quasi-static_2007,oreshkov_adiabatic_2010,avron_adiabatic_2012,venuti_adiabaticity_2016},
and for mixing times of dynamical semigroups
\cite{temme_2-divergence_2010,kastoryano_rapid_2013,temme_lower_2013}.
As we shall see, a naive first attempt to carry out such an estimation
leads to a conundrum: relaxation \emph{seems} to always be faster than
adiabatic preparation. 
However, this pessimistic result (for adiabatic preparation) is based
on a bound for the open-system adiabatic theorem
\cite{venuti_adiabaticity_2016} that essentially mimics the
closed-system result \cite{jansen_bounds_2007}. It represents a
worst-case scenario that ignores the extra structure provided by the
thermalizing dynamics. We resolve the conundrum by estimating the
adiabatic time for the thermalizing case 
in the limit of
zero temperature and show
that in this case, i.e., for sufficiently low  temperatures, adiabatic
preparation can beat thermal relaxation after all.  

We also provide several numerical examples confirming the existence of
both scenarios. Namely, in the case of unstructured (i.e., not
thermal) Lindbladian dynamics, we give an example where the TTSS for
both relaxation and adiabatic preparation is polynomial in the system
size, with relaxation being faster. In contrast, for thermalizing
processes described by a Davies-Lindblad master equation, we give
examples where adiabatic preparation becomes advantageous for
sufficiently low  temperatures and small system-bath coupling. We
expect our conclusions to apply outside of the QA and AQC context,
e.g., for various protocols for faster-than-classical adiabatic
preparation of interesting physical states
\cite{AharonovTa-Shma,Schaller:08,Hamma:2008jk,PhysRevB.91.134303}. 

The structure of this paper is as follows. In Sec.~\ref{sec:2} we
provide the general theoretical framework for relaxation and adiabatic
preparation under Lindbladian evolution. In Sec.~\ref{sec:3} we
analyze adiabatic preparation in the zero temperature limit and
explain why adiabatic preparation can, after all, beat relaxation. In
Sec.~\ref{sec:examples} we analyze three models (dissipative
quasi-free fermions, a single qubit coupled to a thermal bath, and the
\qt{spike} problem coupled to a thermal bath) to test the predictions
of the previous section. We conclude in Sec.~\ref{sec:7}, and provide
additional technical details in the appendix. 

\section{Relaxation and adiabatic preparation under Lindbladian evolution}
\label{sec:2}

The system's Hilbert space $\mathcal{H}$
is assumed to be of finite dimension $d_{\mathcal H}$. Let $H(t)=\sum_{m}E_{m}(t)\ketbra{m(t)}$ be the spectral decomposition of the system Hamiltonian. 
We call the set of instantaneous eigenvectors $\{\ket{m(t)}\}$ of $H(t)$ the energy eigenbasis.
The system's density matrix evolves according to the master equation 
\begin{equation}
\frac{d\rho}{dt}=\mathcal{L}(t)[\rho]\ ,
\end{equation}
where the time-dependent generator $\mathcal{L}(t)$ can be written as $\mathcal{L}(t) = \mathcal{K}(t)+\mathcal{D}(t)$, where the coherent term $\mathcal{K}(t) = -i[H(t),\bullet]$ and where the dissipator $\mathcal{D}(t)$ is in Lindblad
form, for all times $t\in[0,\tau]$. We also assume that the Lindbladian $\mathcal{L}$
is a function of $t/\tau$ such that $1/\tau$ is the (slow) rate
of change ($\tau$ is large). Generators of this form can be derived from realistic
microscopic models provided the decay time  of the correlations
of the reservoir is much shorter than the typical relaxation time
of the system \cite{davies_markovian_1974,gorini_completely_1976}.
In the case where the system Hamiltonian changes very slowly one obtains
the so-called Davies generator with time-dependent Lindblad operators
\cite{alicki_quantum_2007,PhysRevA.73.052311,albash_quantum_2012}.

The goal for both strategies is to prepare the asymptotic steady-state of the final Lindbladian $\mathcal{L}(\tau)$, i.e., the state $\rho_{\mathrm{SS}}(\tau)$ that satisfies 
\beq
\mathcal{L}(\tau)[\rho_{\mathrm{SS}}(\tau)]=0\ .
\eeq
We assume that this steady state is unique. 

For the adiabatic preparation, the system is initialized at $t=0$ in the state $\rho_{\mathrm{SS}}(0)$, the steady state of $\mathcal{L}(0)$, which is assumed to be easy to prepare.  
The time-evolved density matrix is 

\beq
\rho_{\mathrm{adia}}(\tau)=\mathcal{E}_{\tau}[\rho(0)]\ ,
\eeq 
where the evolution operator $\mathcal{E}_{\tau}$ is given by
\begin{equation}
\mathcal{E}_{\tau}=\mathrm{Texp}\left[\int_{0}^{\tau}\mathcal{L}(t)dt\right]\ ,\label{eq:adia_dyn}
\end{equation}
%
and where $\mathrm{T}$ denotes time ordering. 
If the Lindbladian is changed slowly enough, i.e., if $\tau$ is large enough,
the system evolves close to the steady state of $\mathcal{L}(\tau)$.

In the relaxation-based strategy we fix the Lindbladian to its value at $t=\tau$, such that the corresponding relaxation dynamics is given by 
\begin{equation}
\rho_{\mathrm{relax}}(t)=e^{t\mathcal{L}(\tau)}[\rho(0)] \ ,
\label{eq:relax_dyn}
\end{equation}
where $\rho(0)$ is a suitably chosen initial state, e.g., the totally mixed state. 

The TTSS is defined for both the relaxation and the adiabatic preparation as: 
\beq
\tau_{\alpha}  \equiv \min\{t \ | \ d[\rho_{\alpha}(t),\rho_{\mathrm{SS}}(\tau)]\le\epsilon\}\ ,
\quad \alpha\in\{\mathrm{adia},\mathrm{relax}\}
\label{eq:tau_alpha}
\eeq
where $\epsilon$ is a given error target, and $d[\bullet,\bullet]$
is a meaningful distance between density matrices (e.g., the
trace-norm distance). Clearly, $\tau_{\mathrm{adia}}$ and $\tau_{\mathrm{relax}}$ are perfectly comparable in this setting.

Let 
\beq
\mathcal{L}(t)=\sum_{j}\lambda_{j}(t)P_{j}(t)+D_{j}(t)
\eeq 
be the instantaneous Jordan decomposition of $\mathcal{L}(t)$, where the $\lambda_{j}(t)$ are the 
instantaneous eigenvalues, $P_{j}(t)$ are the invariant projectors and $D_{j}(t)$ are the nilpotent terms \cite{kato_perturbation_1995}. Note that it follows from the Lindblad structure that the eigenvalues have the form 
\beq
\lambda_{j}(t)=-\eta_{j}(t)+i\sigma_{j}(t)\ , \quad
\sigma_{j}(t)\in\mathbb{R}\ , \ \ \mathbb{R}\ni\eta_{j}(t)\ge0\ ,
\label{eq:lambdas}
\eeq 
and moreover $\lambda_{0}(t)=0$ and $D_{0}(t)=0$
\cite{alicki_quantum_2007,venuti_adiabaticity_2016}. For convenience
we order the eigenvalues in order of increasing $\eta_j$. 

We proceed to estimate the TTSS in the two approaches. 

\subsection{TTSS for relaxation}
If $\mathcal{L}(\tau)$ is of Davies type (or satisfies a more general {reversibility} condition), it is known that \cite{temme_2-divergence_2010,kastoryano_rapid_2013,kastoryano_quantum_2013}
\begin{equation}
\left\Vert \rho_{\mathrm{relax}}(t)-\rho_{\mathrm{SS}} (\tau)\right\Vert_1 \le\sqrt{\left\Vert \rho_{\mathrm{SS}} (\tau)^{-1}\right\Vert _{\infty}}e^{-t\Delta_{\mathrm{relax}}} \ , 
\label{eq:chi2_bound}
\end{equation}
where $\left\Vert X \right\Vert _{\infty}$ denotes the operator norm
(maximum singular value) of $X$, and the relaxation gap is given by
$\Delta_{\mathrm{relax}}=\min_{j>0}[-\mathrm{Re}(\lambda_{j}(\tau))]=\eta_1(\tau)$
and is assumed to be positive.
The bound~\eqref{eq:chi2_bound} is valid for all possible initial states $\rho_{0}$
and as such is a worst-case scenario. 
 An alternative bound is $\left\Vert \rho_{\mathrm{relax}}(t)-\rho_{\mathrm{SS}}(\tau)\right\Vert _1\le\sqrt{2\ln\left\Vert \rho_{\mathrm{SS}}(\tau)^{-1}\right\Vert _{\infty}}e^{-t\alpha}$.
This has an exponentially improved prefactor at the expense of a smaller
(so called Log-Sobolev) constant: $\alpha\le\Delta_{\mathrm{relax}}$
\cite{kastoryano_rapid_2013,kastoryano_quantum_2013}. 
Note that for particular initial states, the convergence of the
relaxation process may be faster than $\propto
e^{-\Delta_{\mathrm{relax}} t}$, e.g., when
$\rho_0$ does not have a component along $P_1$ (the invariant projector corresponding to the first excited state), since then the relaxation rate is governed by some $\eta_j>\eta_1$. 
If the steady state is thermal then $d_{\mathcal{H}} \le \left\Vert
\rho_{\mathrm{SS}}(\tau)^{-1}\right\Vert _{\infty} \le d_{\mathcal{H}}
\exp[\beta (\Vert H\Vert_\infty - \Vert H^{-1}\Vert_\infty^{-1} )]$, and
so the prefactor in Eq.~\eqref{eq:chi2_bound} is at most exponential
in the total number of sites (or qubits) $n$, provided $H$ is local 
\cite{kastoryano_quantum_2013}.

These
considerations suggest that 
$\left\Vert \rho_{\mathrm{relax}}(t)-\rho_{\mathrm{SS}}(\tau)\right\Vert_1 \sim Ae^{-t\Delta_{\mathrm{relax}}}$
is a reasonable estimate. 
In other words, our estimate for the TTSS, or relaxation time up to an error
$\epsilon$ is 
\begin{equation}
\tau_{\mathrm{relax}}\sim\frac{1}{\Delta_{\mathrm{relax}}}\ln(A/\epsilon)  \equiv \tau^e_{\mathrm{relax}}\ , 
\label{eq:T_relax}
\end{equation}
where the \qt{e} superscript denotes that the right-hand side is an estimate.
As just noted above, based on the result for Davies generators we {expect} that the prefactor $\ln (A)$ is at most polynomial in $n$.

\subsection{TTSS for adiabatic preparation}
Next, we estimate the TTSS in the case of adiabatic preparation. Adiabatic theorems for open systems were proven in Refs.~\cite{joye_general_2007,salem_quasi-static_2007,oreshkov_adiabatic_2010,avron_adiabatic_2012},
but we will need the version of Ref.~\cite{venuti_adiabaticity_2016}, which also gave gap estimates. 
In particular, it was shown there that 
if $\mathcal{L}(t)$ depends smoothly on $t$, and the adiabatic
gap is given by $\Delta_{\mathrm{adia}} = \min_{t\in[0,\tau]}
\min_{j}\left|\lambda_{j}(t)\right|>0$, then
\beq
\left\Vert \rho_{\mathrm{adia}}(\tau)-\rho_{\mathrm{SS}}(\tau)\right\Vert_1 \le B/\tau\ .
\label{eq:adia-SS}
\eeq
The constant $B$ can be taken as 
\begin{multline}
B=\left\Vert S(\tau)\rho_{\mathrm{SS}}'(\tau)\right\Vert_1 +\left\Vert S(0)\rho_{\mathrm{SS}}'(0)\right\Vert_1 \\
 + \frac{1}{\tau}\int_{0}^{\tau} dt \left\Vert S'(t) \rho_{\mathrm{SS}}'(t)+S(t)\rho_{\mathrm{SS}}''(t)\right\Vert_1 \ ,
\label{eq:C-adia}
\end{multline}
where primes denote differentiation with respect to $s=t/\tau$, $S(t)=\lim_{z\to0}Q_{0}(t)[z-\mathcal{L}(t)]^{-1}Q_{0}(t)$ is the reduced resolvent of $\mathcal{L}$, 
$P_0(t)$ denotes the (instantaneous) spectral projection of $\mL(t)$
with eigenvalue zero, and $Q_{0}(t)=\1-P_{0}(t)$. 
As mentioned above,
we assumed that the steady state is unique for all $t$. 
From this bound one thus obtains for the adiabatic TTSS:
\begin{equation}
{\tau}_{\mathrm{adia}}\sim\frac{B}{\epsilon} \equiv \tau^e_{\mathrm{adia}}\ .
\label{eq:T_adia_prediction}
\end{equation}
Thanks to the adiabatic theorem, we are guaranteed that $\tau_{\mathrm{adia}} \le \tau^e_{\mathrm{adia}}$.

Similarly to 
the closed-system case, there is much interest in understanding
the dependence of the constant $B$ on the Lindbladian gap. Roughly, each $S$
term and each time derivative add an inverse power of this
gap. Thus, in general Eq.~\eqref{eq:C-adia} predicts $B\sim O(\Delta_{\mathrm{adia}}^{-3})$, in analogy to the closed-system case \cite{jansen_bounds_2007}. However,
for Davies generators
 $B\sim O(\Delta_{\mathrm{adia}}^{-2})$ \cite{venuti_adiabaticity_2016}. For the sake of generality we write $B \sim C\Delta_{\mathrm{adia}}^{-\psi}$ 
where the constant $C$ can, in principle, be obtained from Eq.~\eqref{eq:C-adia}.
Accordingly, the estimate of the TTSS for adiabatic preparation becomes
\begin{equation}
\tau_{\mathrm{adia}}\sim\frac{C}{\Delta_{\mathrm{adia}}^{\psi}\epsilon}\ .\label{eq:T_adia}
\end{equation}

Regarding the exponent $\psi$, it can be shown that $\psi=1$ for unitary families
of Lindbladians \cite{venuti_adiabaticity_2016}, while it is known
that it can be reduced to $1$ in some cases by a careful choice
of the interpolation schedule between the initial and final Hamiltonians \cite{roland_quantum_2002,PhysRevLett.103.080502}. Accordingly, we shall keep $\psi$ free, but assume that $\psi\ge1$. 

The prefactor $C$ is more subtle. In the closed-system setting it is straightforward to show that for smooth, local, Hamiltonians (and taking $\psi=3$) $C$ is a polynomial in the number of sites $n$ \cite{jansen_bounds_2007,lidar:102106}. One might thus be tempted to conjecture that, similarly, for smooth, local Lindbladians the constant $C$ also scales like a polynomial in $n$. However, this is not generally true.%
\footnote{Obtaining a result analogous to the closed system scaling of $C$ in the open-system setting is much
harder, essentially because the trace-norm is not invariant under the
transformation that diagonalizes the Lindbladian, which means that bounds on $\|S\|$ may hide an additional, non-trivial dependence on $n$.}  
We show in Sec.~\ref{sec:3} that in the zero temperature limit the Hamiltonian gap $\Delta_{1,0} = E_1-E_0$ also plays a role, i.e., that $C \sim C' /\Delta_{1,0}^{\psi'}$, where the new constant $C'$ depends at most polynomially on $n$ (through $\| H\|$), and also $\psi'\geq 1$. For hard computational problems the Hamiltonian gap is typically exponentially small in $n$. We may thus conjecture that, more generally (even at non-zero temperature) a more detailed estimate of the adiabatic TTSS is:
\beq
\tau'_{\mathrm{adia}}\sim\frac{C'}{\Delta_{\mathrm{adia}}^{\psi}\Delta_{1,0}^{\psi'}\epsilon}\ .
\label{eq:T_adia'}
\eeq
We provide a more formal justification for Eq.~\eqref{eq:T_adia'} in Sec.~\ref{sec:3}.

\subsection{Comparing the TTSSs for relaxation and adiabatic preparation}
We are now ready to compare the relaxation and adiabatic approaches, whose TTSSs are captured by Eqs.~\eqref{eq:T_relax} and \eqref{eq:T_adia}, respectively. As we have argued, the relaxation prefactor $\ln A$ is expected to be polynomial in $n$, while we cannot rule out an exponential dependence on $n$ of the prefactor $C$ in Eq.~\eqref{eq:T_adia}. However, for the time being let us ignore this possibility, and focus purely on the effect of the Lindbladian gaps on the TTSSs.

The setting we have in mind is that of a typical hard problem, where we expect the gaps to be exponentially small in $n$. The scaling of the TTSS for the two scenarios
is then determined primarily by their respective Lindbladian gaps, which we recall here for clarity:
\bes
\begin{align}
\Delta_{\mathrm{adia}} &
=\min_{t\in[0,\tau]}\min_{j>0}\sqrt{\sigma_{j}^{2}(t)+\eta_{j}^{2}(t)} \label{eq:adia_gap}\\ 
\Delta_{\mathrm{relax}} & =\min_{j>0}\{\eta_{j}(\tau) \ |\ \eta_{j}(\tau)\neq 0\}\ , \label{eq:relax_gap}
\end{align}
\ees
where $\sigma_{j}$ and $\eta_{j}$ are, respectively, the imaginary and negative real parts of the eigenvalues $\lambda_j(t)$ of the Lindbladian generator, ordered by the real parts [Eq.~\eqref{eq:lambdas}].

We note first that it certainly is possible to formally ensure that $\Delta_{\mathrm{adia}} > \Delta_{\mathrm{relax}}$. For example, assume that $|\lambda_j(t)|>|\lambda_1(t)|$ $\forall j>1$ and $\forall t$, that $\eta_1(t)$ is monotonically decreasing $\forall t$, and that $\sigma_1(t)\neq 0$ $\forall t$. Then $\Delta_{\mathrm{adia}}=\sqrt{\sigma_{1}^{2}(t)+\eta_{1}^{2}(t)} \geq \eta_{1}(\tau) = \Delta_{\mathrm{relax}}$. However, this formal scenario is not physically well motivated.

On the other hand, note that $\Delta_{\mathrm{adia}}  \leq  \sqrt{\sigma_1^2(\tau) + \eta_1^2(\tau)}$, so that:
\begin{claim} 
\label{claim:1}
If $\sigma_1(\tau) = 0$, then necessarily $\Delta_\mathrm{adia} \leq \Delta_{\mathrm{relax}}$.
\end{claim}
Note also that the 
the situation where $\sigma_1(\tau) = 0$
is quite common. For example, for
Davies generators, if the initial state is diagonal in the energy
eigenbasis the dynamics are entirely described by the Pauli master
equation. The Pauli generator is similar to a Hermitian operator%
, which ensures 
that its eigenvalues are real \cite{alicki_quantum_2007}, 
so that indeed
$\sigma_1(\tau) = 0$. 

This is clearly a potential source of concern from the perspective of a speedup via adiabatic preparation. 
Moreover, two other factors also conspire against the adiabatic approach: the fact that $\psi\ge 1$ [Eq.~\eqref{eq:T_adia}], and the exponentially more favorable scaling with the error parameter $\epsilon$ in the relaxation approach.

What if $\sigma_{1}(\tau) \neq 0$? It might seem that one can circumvent the pessimistic conclusion regarding adiabatic preparation by ensuring that $\sigma_{1}(\tau)$ is sufficiently large.
At first sight, an almost trivial strategy to increase $\sigma_{1}(\tau)$ is the following. 
Assume that $[\mathcal{K},\mathcal{D}]=0$. Increasing the magnitude of $\mathcal{K}$ (while keeping $\mathcal{D}$ fixed) results in eigenvalues with a larger
imaginary part. This change has no effect on the relaxation
process, but it may have an effect on the adiabatic one, as the Lindbladian 
is now more dominated by the coherent term and hence is \qt{more quantum}.
In other words, a simple way to obtain a large value of $\left|\sigma_{1}(\tau)\right|$ is to increase the magnitude of the coherent term.
This is, in fact, precisely the idea used to protect AQC using dynamical decoupling \cite{PhysRevLett.100.160506}, and is more generally a commonly utilized strategy to enact
quantum information primitives in the presence of a dissipative environment. 

However, note that in our context this strategy is only effective up to a point. To see this, 
assume that we rescale the coherent term as $\mK\to \al \mK$, where $\al>1$. 
The eigenvalues in Eq.~\eqref{eq:lambdas} become $\lambda_j(\al) = - \eta_j +i \al \sigma_j$. Then $|\lambda_j(\al)|$ increases as $\al$ increases, except for those $\lambda_j$ for which $\sigma_j =0$ (we refer to such eigenvalues as belonging to the $\mK=0$ sector). Such eigenvalues always exist, and the corresponding eigenstates are also eigenstates of $\mL$.%
\footnote{Consider a state that is diagonal in the energy eigenbasis, i.e., $\rho = \sum_m a_m \ketbra{m}$. Since $\ket{m}$ is an eigenvector of $H$, it follows that $\ketbra{m}$ is an eigenstate of $\mK = -i[H,\bullet]$ with eigenvalue zero, and hence $\rho$ belongs to the nullspace of $\mK$. Recall that $\mathcal{L}=\mathcal{K}+\mathcal{D}$ and that the eigenvalues of $\mathcal{L}$ [Eq.~\eqref{eq:lambdas}] are $\lambda_j = - \eta_j +i \sigma_j$, where $i \sigma_j$ is an eigenvalue of $\mK$ and $\eta_j$ is an eigenvalue of $\mD$.  
Since by assumption $[\mK,\mD]=0$, the zero-eigenvalue eigenstates of $\mK$ are shared by $\mD$, and hence are also eigenstates of $\mL$.}
Now suppose $|\lambda_1|$, with $\sigma_1\neq 0$, is the smallest eigenvalue, i.e., it determines $\Delta_{\mathrm{adia}}$. Then, as $\al$ increases, at some point $|\lambda_1(\al)|$ will become larger than the modulus of one of the $\mK=0$ sector eigenvalues, since these are unaffected by increasing $\al$. Therefore, for sufficiently large $\al$, one of the $\mK=0$ sector eigenvalues becomes the eigenvalue with the smallest modulus, and determines $\Delta_{\mathrm{adia}}$. At this level-crossing point there is no further advantage to increasing $\al$.

Finally, let us return to the effect of the prefactor $C$ in Eq.~\eqref{eq:T_adia}, which is captured by Eq.~\eqref{eq:T_adia'}.  
Comparing the latter to Eq.~\eqref{eq:T_relax}, it is clear that matters only become worse for the adiabatic preparation procedure relative to relaxation. Namely, in Eq.~\eqref{eq:T_adia'}, where $C'$ is polynomial in $n$, the extra Hamiltonian gap factor in the denominator will cause $\tau'_{\mathrm{adia}}$ to acquire another exponential factor in $n$ for hard problems. 

We are thus left with the pessimistic conclusion that apparently adiabatic 
evolution
is, quite generally, inferior to relaxation
when the goal is steady state preparation. 
For those cases where thermal relaxation can be performed via a classical algorithm, such as Davies generators with a classical final Hamiltonian (the setting of all QA optimization problems), this would also mean that adiabatic quantum preparation is slower than classical algorithms. Given that in reality the open system setting is unavoidable, this would appear to dash all hopes for an experimental quantum speedup via QA, and in particular would appear to doom experimental efforts at realizing physical quantum annealers \cite{Dwave}. 

However, we shall next see that this conclusion is, in fact, premature and overly pessimistic. The basic reason is the fact that, having been derived as 
a bound,
Eq.~\eqref{eq:T_adia} represents a worst case scenario. 
In fact, Lindblad master equations (may) have
considerably more structure than is captured by 
such
bounds, as is revealed by a careful study of the zero-temperature limit in the next section, and by numerical examples in Sec.~\ref{sec:examples}. In fact, we shall encounter an example (the \qt{spike}) for which despite $\sigma_1(\tau) = 0$ being satisfied, adiabatic preparation will turn out to best relaxation. 

\section{Adiabatic preparation in the zero temperature limit and why adiabatic preparation can, after all, beat relaxation}
\label{sec:3}

In this section we give a more precise estimate of $\tau_{\mathrm{adia}}$ in the low temperature limit. Specifically, our aim is to estimate the terms in Eq.~\eqref{eq:C-adia} in this limit. To this end we assume that the generator is thermalizing (e.g., of Davies type)
with a unique steady state. In this case, the steady state is  given
by $\rho_{\mathrm{SS}}(s)=e^{-H(s)/T}/Z$ with $Z=\mathrm{Tr}e^{-H(s)/T}$. Again, let
$H=\sum_{m}E_{m}|m\rangle\langle m|$ 
be the spectral decomposition of $H$ (we assume a non-degenerate
spectrum and omit the $s$ dependence for notational simplicity). One also has
$\mathcal{L}[|l\rangle\langle m|] = \lambda_{l,m} |l\rangle\langle m|$
for $m\neq l$ with 
\beq
\lambda_{l,m}=-i\Delta_{l,m}- \eta_{l,m}\ , \quad \Delta_{l,m}=E_{l} -E_m\ .
\eeq 
The positive numbers $\eta_{l,m}$ can also be obtained directly from the
generator $\mathcal{L}$ \cite{davies_markovian_1974}.
Since by assumption the
Hamiltonian ground-state is non-degenerate, we have
$\lim_{T\to0}\rho_{\mathrm{SS}}=|0\rangle\langle0|$. We then obtain (see Appendix~\ref{app:zeroT} for details): 
\begin{equation}
\lim_{T\to0}S\rho_{\mathrm{SS}}'=-\sum_{l\neq0}\frac{\langle l|H'|0\rangle}{\Delta_{l,0}}\frac{1}{\lambda_{l,0}}|l\rangle\langle0|+\hc
\end{equation}
 This corresponds to a trace-norm contribution of
\bes
\begin{align}
\label{eq:19a}
\lim_{T\to0}\left\Vert S\rho_{\mathrm{SS}}'\right\Vert _{1}  &=  2\sqrt{\sum_{l\neq0}\left|\frac{\langle l|H'|0\rangle}{\Delta_{l,0}\lambda_{l,0}}\right|^{2}}\\
 & \approx  2\left|\frac{\langle 1|H'|0\rangle}{\Delta_{1,0}\lambda_{1,0}}\right|\ , \label{eq:Srhop}
\end{align}
\ees
where in the last line we approximated the sum by its leading term, which we assumed to be at $l=1$. Since $|\lambda_{1,0}| \geq \Delta_{1,0}$, this assumption is justified when the expression in Eq.~\eqref{eq:19a} is evaluated at $s_{\min}$, i.e., $s$ such that the \emph{Hamiltonian} gap $\Delta_{1,0}$ is minimum. 

The other terms in Eq.~\eqref{eq:C-adia} can be obtained in a similar, though lengthier way (see Appendix~\ref{app:zeroT}).  It turns out that as $T\to 0$,
$S'\rho_{\mathrm{SS}}'+S\rho_{\mathrm{SS}}''$ becomes a rank-four
operator. In evaluating its trace-norm we only keep the leading contributions. The result is:
\begin{align}
\varepsilon(t) &:= \left\Vert S\rho''+S'\rho'\right\Vert _{1}  \nonumber \\
& \approx \Bigg[ 4\frac{\left|\langle1|H'|0\rangle\right|}{\Delta_{1,0}^{2}}\left|\frac{\langle1|H'|1\rangle-\langle0|H'|0\rangle}{\lambda_{1,0}}\right| \nonumber \\
& \qquad +4\left|\frac{\langle1|H'|0\rangle}{\Delta_{1,0}}\right|^{2}\left|\mathrm{Re}
  \left(\frac{1}{\lambda_{1,0}}\right)\right| \Bigg].
  \label{eq:eps(t)}
\end{align}
Note that $|\mathrm{Re}(1/\lambda_{1,0})|\le 1/|\lambda_{1,0}|\le 1/\Delta_{1,0}$, so that $\varepsilon(t)$ is dominated by the Hamiltonian gap $\Delta_{1,0}$; the bath enters only via the \emph{off-diagonal} Lindbladian eigenvalue $\lambda_{1,0}$.\footnote{We shall see in the single-qubit example considered in Sec.~\ref{sec:5} that this is an important point; since $\Delta_{\mathrm{adia}}$ does contain the diagonal $\lambda_{1,1}$ eigenvalue, which is purely real and is proportional to the system-bath coupling constant $g$, $\Delta_{\mathrm{adia}}$ gives the wrong estimate for $\tau_{\mathrm{adia}}$ in the small $g$ limit, while $\varepsilon(t)$ gives the right estimate.} Moreover, due to the higher power of the Hamiltonian gap it contains, $\varepsilon(t)$ is dominant with respect to Eq.~\eqref{eq:Srhop}, and so the overall contribution to the constant $B$ at zero temperature becomes:
\bes
\label{eq:B_T0}
\begin{align}
\lim_{T\to0} B & \approx \frac{1}{\tau} \int_0^\tau dt\ \varepsilon(t) \label{eq:B_T0a} \\
& \le  \max_{t\in[0,\tau]} \varepsilon(t)
\label{eq:B_T0b}
\end{align}
\ees

The corresponding estimate for the adiabatic time is again $\tau^e_{\mathrm{adia}} \equiv B/\epsilon$. Thus, while Eq.~\eqref{eq:T_adia} contains the minimum Lindbladian gap $\Delta_{\mathrm{adia}}$, Eq.~\eqref{eq:B_T0b} does not, since the time at which $\varepsilon(t)$ is evaluated is determined by the Hamiltonian gap $\Delta_{1,0}$ in Eq.~\eqref{eq:eps(t)}. This implies that $\tau^e_{\mathrm{adia}}$ \emph{can} in principle be
smaller than $\tau_{\mathrm{relax}}$ [Eq.~\eqref{eq:T_relax}]. 

We can now see the explicit justification for Eq.~\eqref{eq:T_adia'}. The expectation that Eq.~\eqref{eq:T_adia} holds with $C$ scaling polynomially in $n$ cannot be fulfilled in the regime $T\ll \min_{t\in[0,\tau]} \Delta_{1,0}$. Namely, even if for smooth local Hamiltonians $|\langle l|
H'|m\rangle | \le \Vert H' \Vert = \mathrm{poly}(n)$, Eq.~\eqref{eq:eps(t)} shows there is a
potential exponential dependence on $n$ due 
to
the Hamiltonian gap. 

Armed with these results, we are now finally able to begin to dispel the pessimistic conclusion about the inferior performance of adiabatic preparation relative to relaxation: in the $T\to 0$ limit, the relevant gap need \emph{not} be the minimum Lindbladian gap but rather the minimum Hamiltonian gap, and thus Eqs.~\eqref{eq:T_relax} and \eqref{eq:T_adia} are, in fact, not always comparable.

How does the result extend beyond $T\to 0$? It can be shown (see Appendix \ref{app:finiteT}) that the leading order positive temperature corrections to Eq.~\eqref{eq:B_T0} are $O\left[T^{-2} \exp{(-\Delta_{1,0}^{\mathrm{min}}/T)}\right]$ where $\Delta_{1,0}^{\mathrm{min}} = \min_{t\in[0,\tau]} \Delta_{1,0}(t)$. Consequently,
Eq.~\eqref{eq:B_T0} is continuous 
as
$T\to0^{+}$.\footnote{Strictly, we assumed a gap condition here, i.e., $\Delta_{1,0}^{\mathrm{min}}>0$. Showing continuity in the case of level crossing can be done using the method of Ref.~\cite{kato_adiabatic_1950}. In any case this formula may drastically underestimate the region of validity of the zero-temperature result. Further studies are needed to correctly address this important point, i.e., what is the range of temperatures such that a speedup present at $T=0$ survives.}
While we only showed that continuity holds for the bound, and not for $\tau_{\mathrm{adia}}$ itself, this does suggest that if $\tau_{\mathrm{adia}} < \tau_{\mathrm{relax}}$ at zero temperature,
the same result will still hold for sufficiently low temperatures. The examples we present in the next section confirm that the situation is indeed more subtle than suggested by the pessimistic conclusions of Sec.~\ref{sec:2}.

\section{Examples}
\label{sec:examples}

In order to check the arguments presented above, we performed extensive numerical
simulations on three models: dissipative quasi-free fermions, a single
qubit coupled to a thermal bath, and the \qt{spike problem} with a
thermal bath. In the first, relaxation beats adiabatic preparation,
while in the second and third example both 
this scenario and its converse
are realized.

\subsection{Dissipative quasi-free fermions}
\label{sec:4}

\begin{figure}[t]
\subfigure[]{\includegraphics[width=4.25cm]{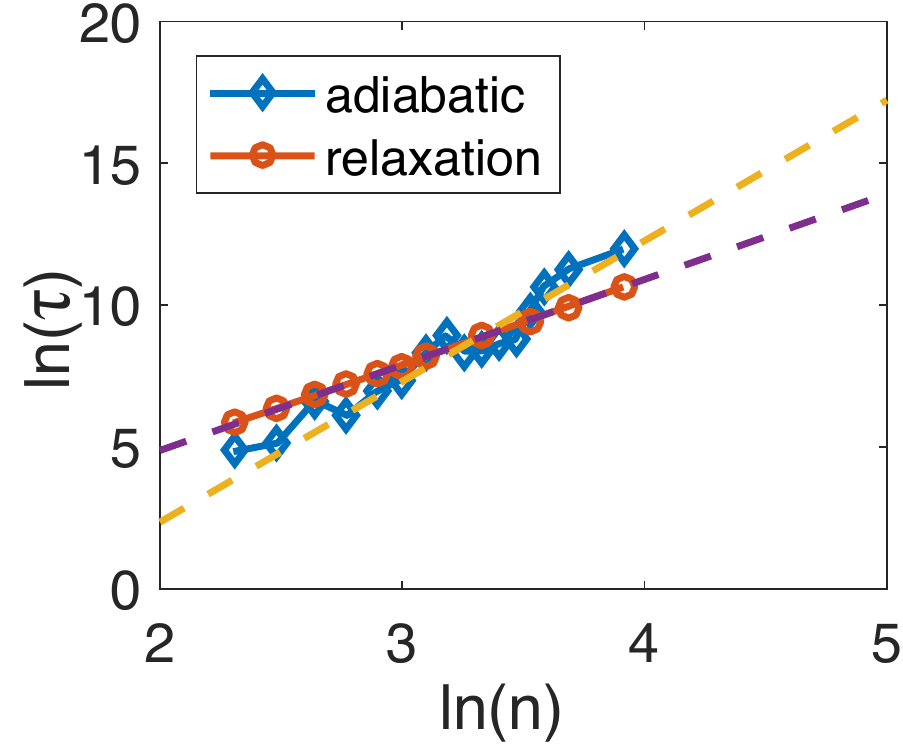}\label{fig:1a}}
\subfigure[]{\includegraphics[width=4.25cm]{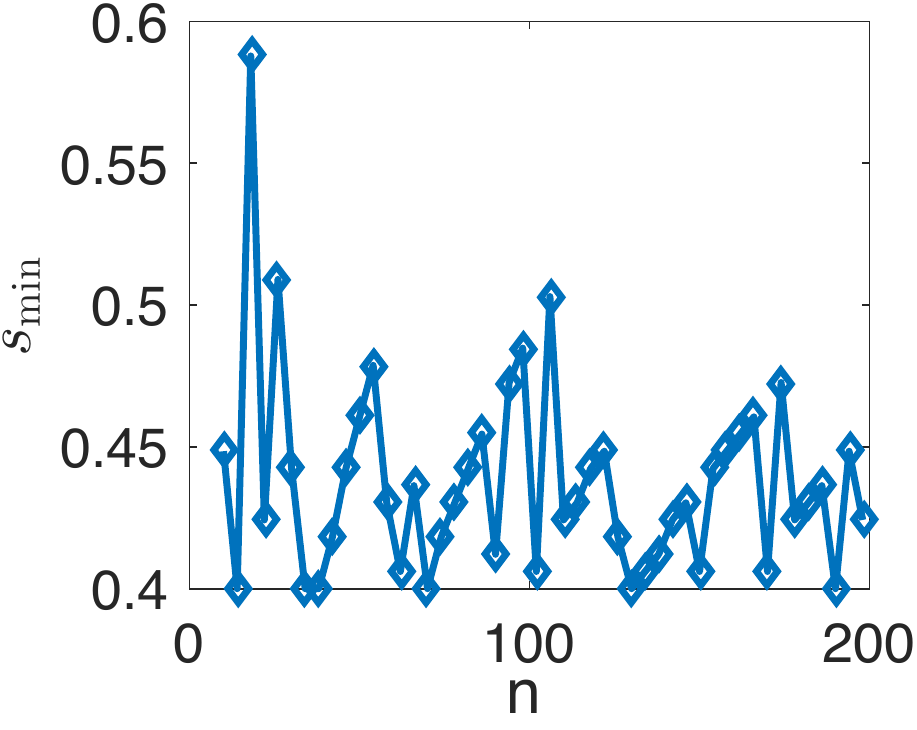}\label{fig:1b}}
\subfigure[]{\includegraphics[width=4.25cm]{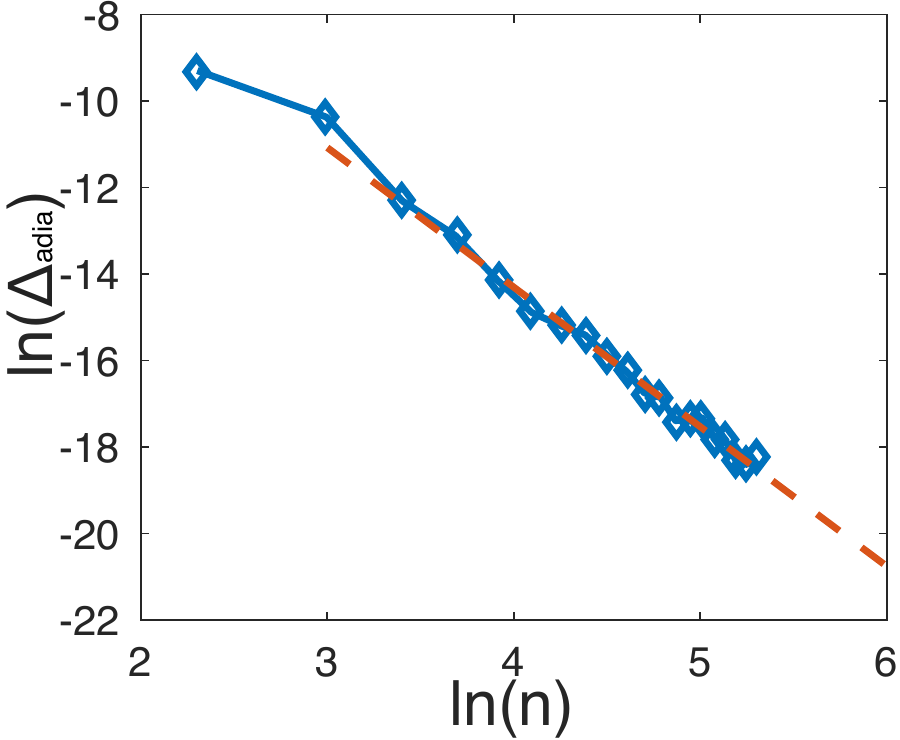}\label{fig:1c}}
\subfigure[]{\includegraphics[width=4.25cm]{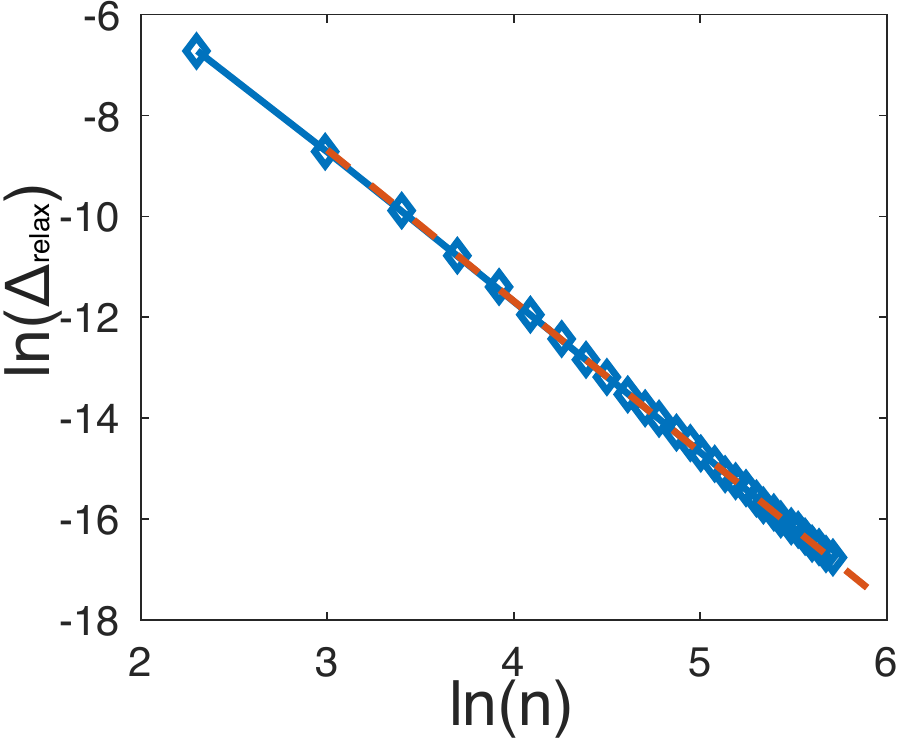}\label{fig:1d}}
\caption{(Color online) Adiabatic \textit{vs}. relaxation-based quantum state preparation
in a quasi-free dissipative system. Gaps are in units of $J$ while times are in units of $J^{-1}$. (a) TTSS
as a function of system size. The fit gives $\tau_{\mathrm{relax}}\sim n^{3.016}$
for the relaxation in accordance with $\tau\sim1/\Delta_{\mathrm{relax}}$.
For the adiabatic case the exponent ranges between $4.4$ using all
$16$ sizes (from $10$ to $60$), to $6.5$ using the largest $5$ sizes.
(b) location of the minimum gap as a function of system size. Lower
panels: adiabatic gap $\Delta_{\mathrm{adia}}$ (c) and relaxation gap $\Delta_{\mathrm{relax}}$ (d) as a function of system size. Both fits give 
$\Delta\sim n^{-3}$ in accordance with \cite{prosen_quantum_2008}.}
\label{fig:Adiabatic-vs-relaxation}
\end{figure}

We first consider a master equation which does not have the extra structure of describing a thermalization process.
The asymptotic steady states are therefore given by non-equilibrium steady states as opposed to thermal states. 

We consider an integrable, quasi-free, fermionic model
with dissipation. This is a time-dependent version of the model considered
in Refs.~\cite{prosen_third_2008,prosen_quantum_2008}, with the
Hamiltonian $H(t)=(1-t/\tau)H_{0}+(t/\tau)H_{1}$, where  
\bes
\begin{align}
H_{0} & =  B\sum_{i=1}^{n}\sigma_{i}^{z}\\
H_{1} & =  J\sum_{i=1}^{n}\left[\frac{\left(1+\gamma\right)}{2}\sigma_{i}^{x}\sigma_{i+1}^{x}+\frac{\left(1-\gamma\right)}{2}\sigma_{i}^{y}\sigma_{i+1}^{y}\right]\ .
\end{align}
\ees
The dissipation consists of placing the two end-spins (at $i = 1, n$) in thermal contact with an external reservoir. Specifically, the Lindblad
operators are given by 
\bes
\begin{align}
L_{1}&=\sqrt{2\Gamma_{1}}\sigma_{1}^{+},\, L_{2}=\sqrt{2\Gamma_{2}}\sigma_{1}^{-}\\
L_{3}&=\sqrt{2\Gamma_{3}}\sigma_{n}^{+},\,L_{4}=\sqrt{2\Gamma_{3}}\sigma_{n}^{-}
\end{align}
\ees
with $\sigma_{j}^{\pm}=(\sigma_{j}^{x}\pm i\sigma_{j}^{y})/2$. The
Lindbladian is defined by 
\begin{equation}
\mathcal{L}(t)[\rho]=-i[H(t),\rho]+\sum_{a=1}^{4} \left[ L_{a}\rho
L_{a}^{\dagger}- \frac{1}{2}\left\{ L_{a}^{\dagger}L_{a},\rho\right\} \right]\ .
\end{equation}
The corresponding evolution maps Gaussian states into Gaussian states.
These can in turn be conveniently characterized in terms of their
covariance matrix. The Lindblad master equation translates into a
differential equation for the covariance matrix (see Ref.~\cite{horstmann_noise-driven_2013}
for details). We use the Bures distance between states for $d$ in Eq.~\eqref{eq:tau_alpha}, 
which can be efficiently computed in terms of
covariance matrices for Gaussian states \cite{banchi_quantum_2014}. As initial state we choose the fully mixed state $\rho_{0}=\1/d$ for the relaxation process,  and the steady state at $t=0$ for the adiabatic process. In order
to determine the TTSS, we solve Eq.~\eqref{eq:tau_alpha} numerically
with a root finder algorithm based on the secant method, where
$\rho_{\al}(t)$ is computed using the Lindblad equation. We
arbitrarily fix $\epsilon=0.1$.  

Our numerical results are displayed in Fig.~\ref{fig:Adiabatic-vs-relaxation}.
Figure~\ref{fig:1a} shows the scaling of the TTSS with system size for both adiabatic preparation and relaxation. Figure~\ref{fig:1b} shows the position of the minimum gap as a function of system size for the adiabatic preparation, which appears irregular, but occurs towards the middle of the evolution. Also shown are the gaps as a function of system size, for both adiabatic preparation [Fig.~\ref{fig:1c}] and relaxation [Fig.~\ref{fig:1d}]; both are consistent with the predicted scaling as $n^{-3}$ \cite{prosen_quantum_2008}.  From this we conclude that $\tau_{\mathrm{relax}}\sim\Delta_{\mathrm{relax}}^{-1}$
and $\tau_{\mathrm{adia}}\sim\Delta_{\mathrm{adia}}^{-\psi}$ where
the exponent $\psi$ ranges between $1.5$ and $2.2$. 
While Fig.~\ref{fig:1a} shows that adiabatic preparation is more efficient at small system sizes than relaxation, its worse scaling with system size causes adiabatic preparation to become less efficient than relaxation for sufficiently large system sizes. 

To conclude, for this model of dissipative quasi-free fermions we find that relaxation is more efficient than the adiabatic process at preparing the steady state of the corresponding Lindbladian.

\subsection{Single qubit coupled to a thermal bath}
\label{sec:5}

Next, we study a single qubit coupled to a thermal bath at inverse temperature $\beta=1/T$. We choose the system Hamiltonian as 
\beq
H(t)=\omega_{x}(1-{t}/{\tau})\sigma^{x}+\omega_{z} ({t}/{\tau}) \sigma^{z}\ ,
\eeq 
whose instantaneous gap is $\delta(t) \equiv \Delta_{1,0}(t)  = 2[(1 - \frac{t}{\tau})^2\omega_x^2  + (\frac{t}{\tau})^2 \omega_z^2]^{1/2}$. We assume that the Lindbladian has the Davies form, which guarantees convergence to the thermal Gibbs state:
\begin{align}
\mathcal L(t) [\rho] = &-i [H(t), \rho] + \sum_{\omega = {0, \pm \delta(t)}} \gamma(\omega) [ L_{\omega}(t) \rho L_{\omega}^{\dagger}(t) 
 \nonumber \\
&  - \frac{1}{2} \left\{ L_{\omega}^{\dagger}(t) L_{\omega}(t), \rho \right\} ] \ .
\label{eq:Lind}
\end{align}
For simplicity we set the Lamb shift Hamiltonian to zero. The positive function $\gamma(\omega)$ encodes the spectral function of the bath,  and satisfies the Kubo-Martin-Schwinger condition $\gamma(-\omega) = e^{-\beta\omega} \gamma(\omega)$ for $\omega>0$ \cite{Kossakowski:1977dk}. We take it to have the Ohmic form 
 \beq
 \gamma(\omega) = \frac{2 \pi g^2 \omega}{1 - e^{-\beta \omega}}\ ,
 \label{eq:gamma}
 \eeq 
where $g$ is the system-bath coupling in the system-bath interaction Hamiltonian, which we assume to have the simple form $H_{SB} = g A\otimes B$, where $A=\sigma^y$.  This choice ensure that the minimum Lindbladian gap
is non-zero throughout the entire evolution (no level crossing). The Lindblad operators are constructed as  $L(\omega)=\sum_{E_{l}-E_m=\omega}{\Pi_m A\Pi_{l}}$, where $\Pi_m$ denotes the orthogonal projection onto the $H$-eigensubspace with energy $E_m$ \cite{Breuer:book}.  We note that while the choice in Eq.~\eqref{eq:gamma} is convenient, it does not capture many important noise sources such as $1/f$ noise (see, e.g., Ref.~\cite{PhysRevLett.118.057703}).

Writing
the Lindbladian in the instantaneous eigenbasis $\{\ket{0(t)},\ket{1(t)}\}$, of $H(t)$, 
one readily
finds two off-diagonal eigenvectors $|m\rangle\langle l|$ (with $m\neq l$ and dropping the time dependence when not strictly needed) with
eigenvalues $\lambda_{1,0}$ and $\lambda_{0,1}=\lambda_{1,0}^*$
with 
$\lambda_{1,0}=-\Gamma-i\delta$,
where $2\Gamma=\gamma(\delta)(1+e^{-\beta\delta})$. In addition there are diagonal eigenvectors $|m\rangle\langle m|$ (which evolve according to the Pauli equation generator), whose eigenvalues are $\lambda_{0,0}=0$ and $\lambda_{1,1}=-2\Gamma$ \cite{venuti_adiabaticity_2016}. The adiabatic gap [Eq.~\eqref{eq:adia_gap}] is defined as the minimum over all non-zero Lindbladian eigenvalues, i.e., letting $s=t/\tau$, here we have
\beq
\Delta_{\mathrm{adia}} = \min_s\{ |\lambda_{1,0}|,|\lambda_{1,1}|\} = \min_s \{|\Gamma+i\delta|,2|\Gamma|\}\ ,
\label{eq:delta_adia_1q}
\eeq
where all quantities are $s$-dependent. We also define
\beq
\Delta_{\mathrm{relevant}} \equiv \min_s\{ |\lambda_{1,0}|\} = \min_s \{|\Gamma+i\delta|\}\ .
\eeq
a quantity we will use momentarily.

We compute the adiabatic and relaxation TTSSs using the trace-norm distance (TND) for $d$ in Eq.~\eqref{eq:tau_alpha}, i.e., we use $d(\rho_1,\rho_2)\equiv \frac{1}{2}\|\rho_1-\rho_2\|_1$.

\begin{figure}
\subfigure[]{\includegraphics[width=4.25cm]{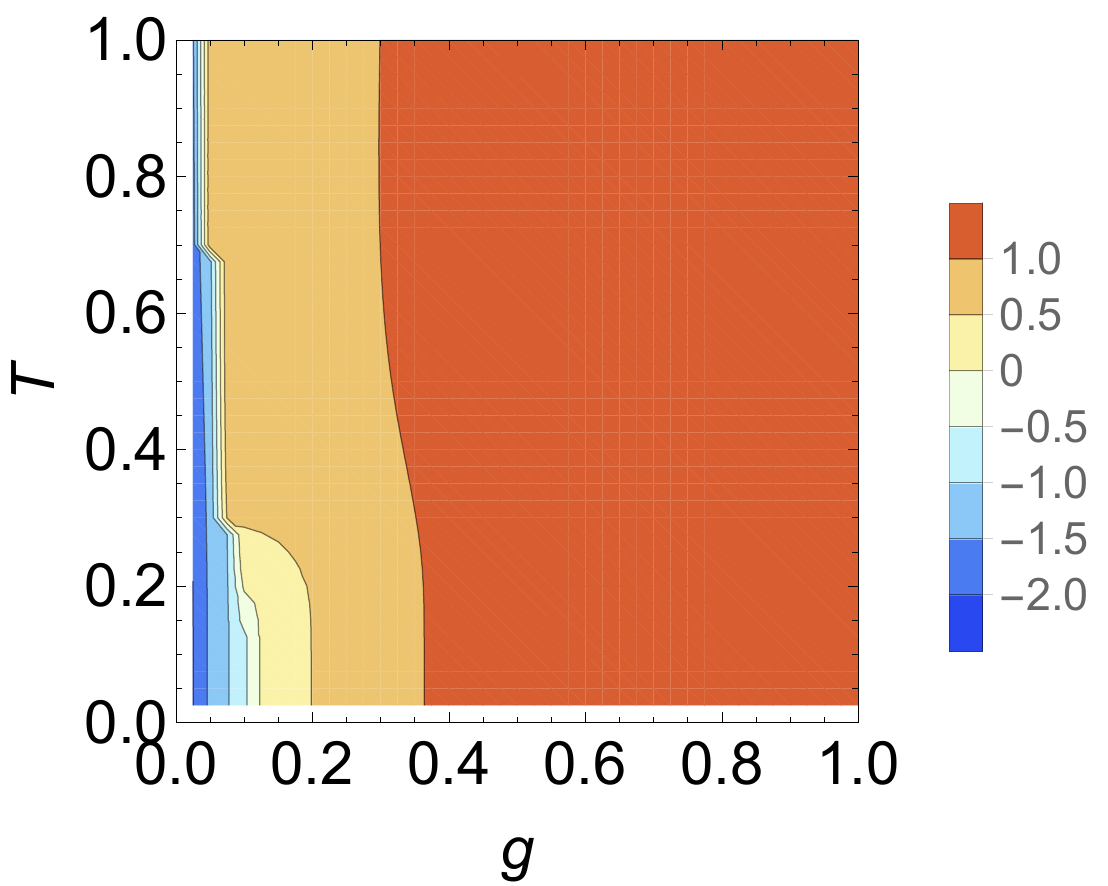}\label{fig:1Qubit-a}}
\subfigure[]{\includegraphics[width=4.25cm]{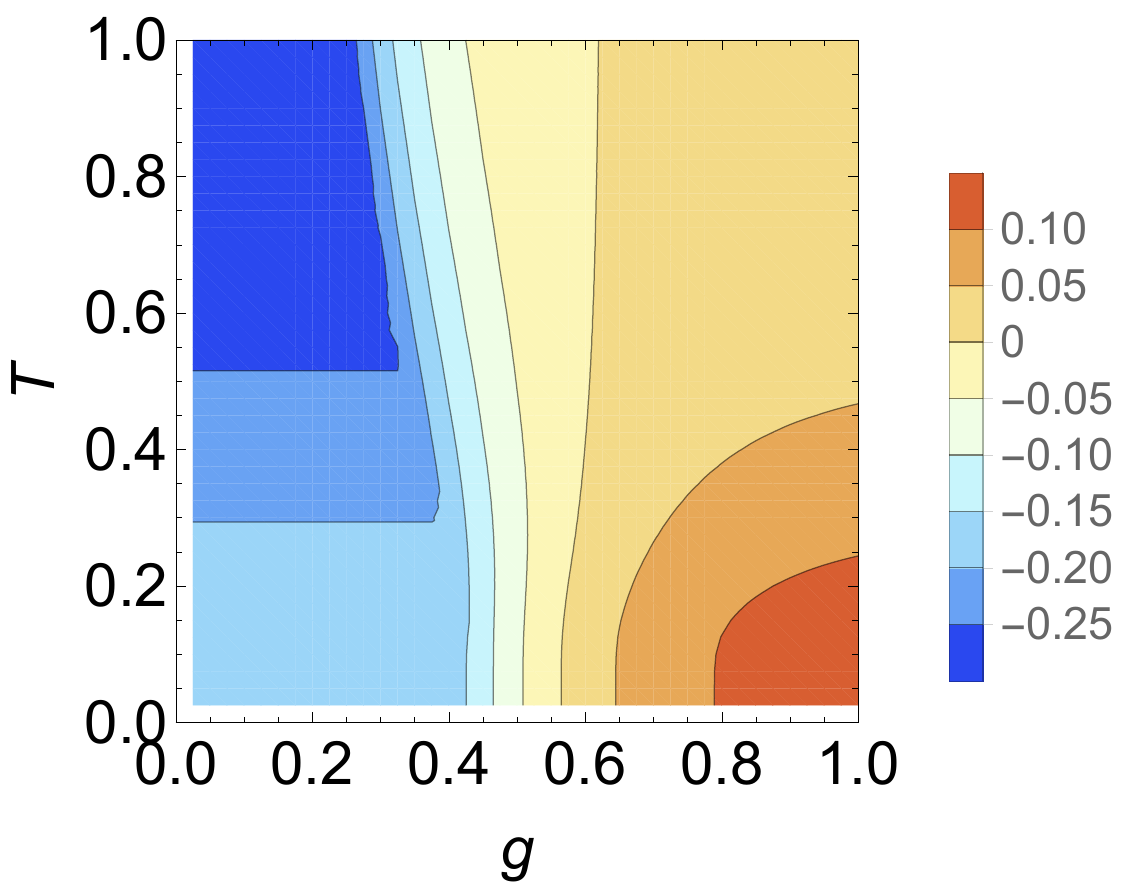}\label{fig:1Qubit-b}}
\caption{{(Color online) (a) The $\log_{10}$ of the ratio $\tau_{\mathrm{adia}}/\tau_{\mathrm{relax}}$
in the $(g,T)$ plane for $\epsilon = 10^{-2}$. Blue indicates that adiabatic preparation is faster, red that relaxation is faster. (b) the $\log_{10}$ of the ratio $\Delta_{\mathrm{relax}}/\Delta_{\mathrm{adia}}$, which roughly correlates as expected with $\tau_{\mathrm{adia}}/\tau_{\mathrm{relax}}$.} 
The Hamiltonian minimum gap $\delta_{\min} = 1$.  Both $T$ and $g$ are measured in units of $\delta_{\min}$. 
}
\label{fig:1Qubit}
\end{figure}

Figure~\ref{fig:1Qubit} presents our numerical results comparing adiabatic preparation and relaxation, in terms of the ratios $\tau_{\mathrm{adia}}/\tau_{\mathrm{relax}}$ [Fig.~\ref{fig:1Qubit-a}] and $\Delta_{\mathrm{relax}}/\Delta_{\mathrm{adia}}$ [Fig.~\ref{fig:1Qubit-b}] in the $(g,T)$ plane.
We observe a rough qualitative agreement between the two ratios, indicating that 
$\psi\approx 1$ [Eq.~\eqref{eq:T_adia}]. Both ratios indicate
that the performance of adiabatic preparation improves as the system-bath
coupling strength decreases, but this effect shrinks as the temperature increases.

\begin{figure}
\subfigure[]{\includegraphics[width=4.15cm]{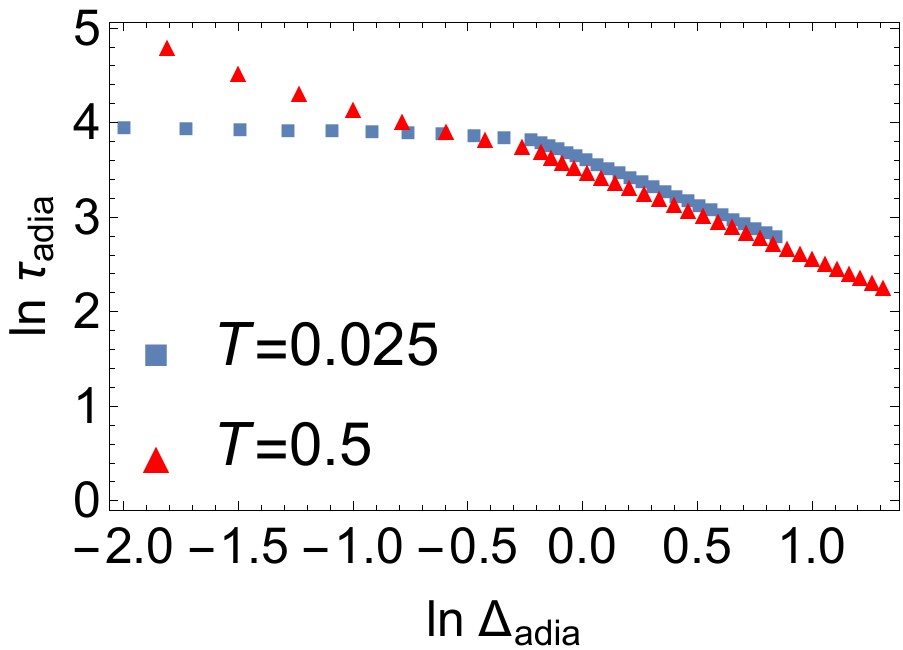}\label{fig:1Qubit-relevant-a}}
\subfigure[]{\includegraphics[width=4.35cm]{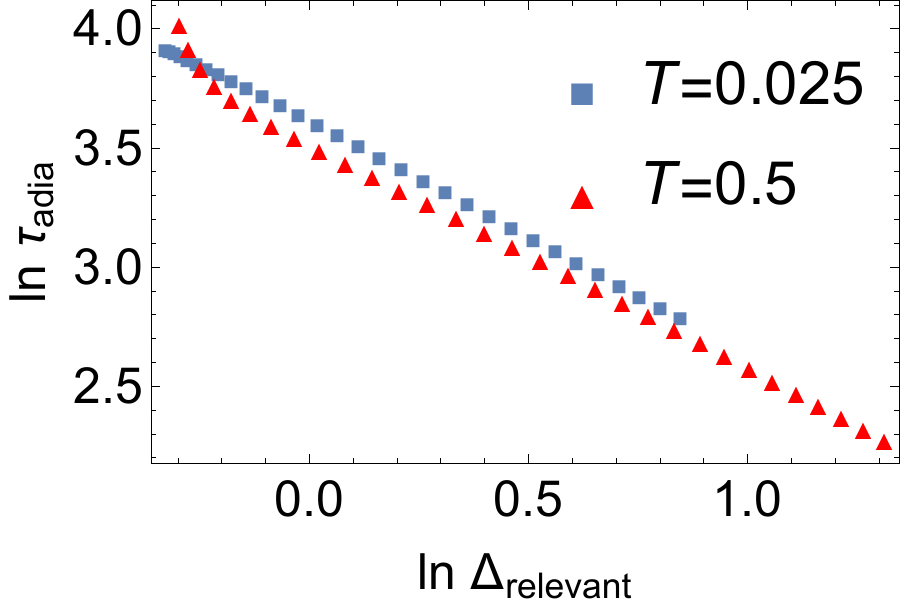}\label{fig:1Qubit-relevant-b}}
\caption{(Color online) (a)  $\ln(\tau_{\mathrm{adia}})$ as a function $\ln(\Delta_{\mathrm{adia}})$ for $\epsilon = 10^{-2}$. The slope to the (left) right of the kink is $-0.85$ ($-0.95$) from a linear fit. (b) $\ln(\tau_{\mathrm{adia}})$ as a function $\ln(\Delta_{\mathrm{relevant}})$ for $\epsilon = 10^{-2}$. The slope for the $T=0.025$ case is $-0.975$ from a linear fit. Here $\tau_{\mathrm{adia}}$ is measured in units of $\delta_\mathrm{min}^{-1}$ and $\Delta_{\mathrm{relevant}}$ is measured in units of $\delta_{\mathrm{min}}$, where $\delta_{\mathrm{min}}$ is the minimum Hamiltonian gap.} 
\label{fig:1Qubit-relevant}
\end{figure}

\begin{figure}
\subfigure[]{\includegraphics[width=4.25cm]{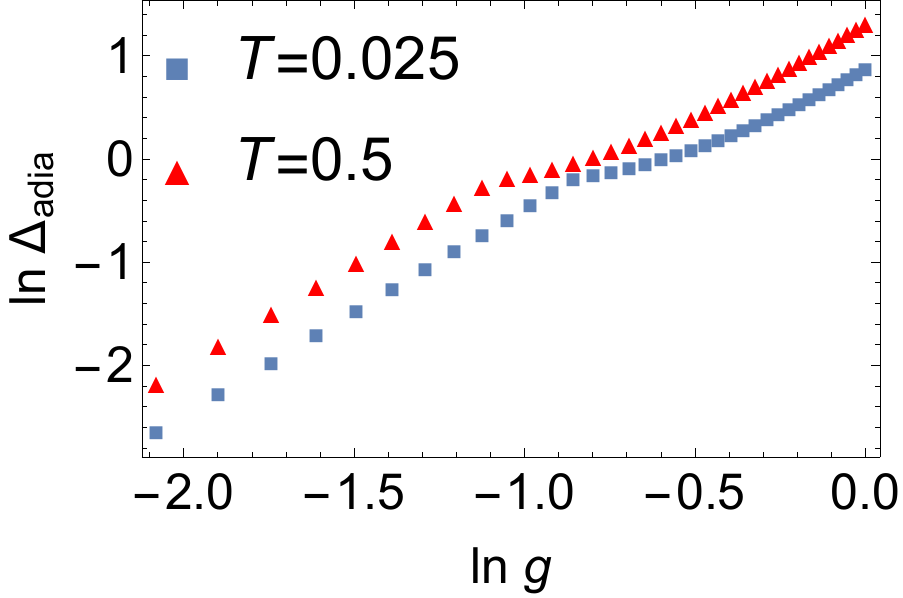}\label{fig:1Qubit-as-func-of-g-a}}
\subfigure[]{\includegraphics[width=4.25cm]{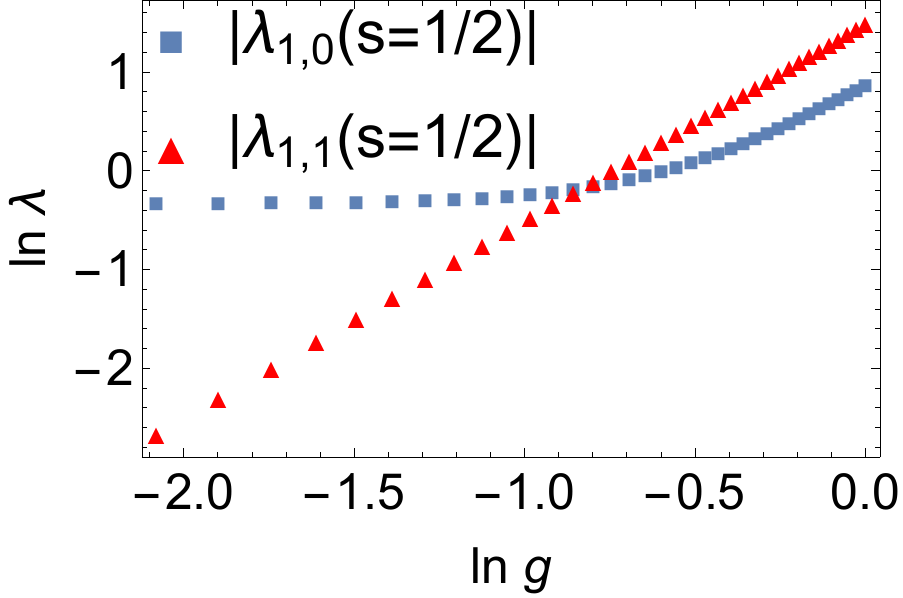}\label{fig:1Qubit-as-func-of-g-b}}
\caption{{(Color online)
 (a) $\Delta_{\mathrm{adia}}$ as a function of the system-bath coupling strength $g$, for two different temperatures. (b) Modulus of the eigenvalues $\lambda_{1,0}$ and $\lambda_{1,1}$ at their respective minimum points ($s=1/2$), as a function of coupling $g$, for $T=0.025$. Their crossing explains the kink seen in (a).  The quantities $\Delta_{\mathrm{adia}}$, $\lambda$ and $g$ are measured in units of the minimum Hamiltonian gap $\delta_{\mathrm{min}}$.} 
}
\label{fig:1Qubit-as-func-of-g}
\end{figure}

To more carefully test the accuracy of the prediction of Eq.~\eqref{eq:T_adia}, we plot $\ln(\tau_{\mathrm{adia}})$ as a function $\ln(\Delta_{\mathrm{adia}})$ in Fig.~\ref{fig:1Qubit-relevant-a}. 
Whereas Eq.~\eqref{eq:T_adia} predicts a linear relation with a  slope of $-\psi$, which holds with $\psi\approx1$ for the high temperature case ($T=0.5$), this clearly  breaks down in the low temperature case ($T=0.025$), where the slope of the $\tau_{\mathrm{adia}}$ data points suddenly bends at $\ln(\Delta_{\mathrm{adia}})\approx 0$.
Indeed, as we argued in Sec.~\ref{sec:3}, in the $T\ll \delta_{\min}$ region
[the Hamiltonian gap $\delta_{\min} = \min_t \delta(t)=1$ in Fig.~\ref{fig:1Qubit-relevant}], 
one should use $\tau_\mathrm{adia} = B/\epsilon$,
with $B$ given by Eq.~\eqref{eq:B_T0b}.  
Since neither the Hamiltonian gap nor the matrix elements of $H'$ 
depend on $g$ and $T$, they can be ignored in this context. 
Hence, according to Eq.~\eqref{eq:B_T0b}, for temperatures $T\ll\delta_{\min}$, 
\beq
\tau_{\mathrm{adia}}\approx \Delta_{\mathrm{relevant}}^{-1}\ , 
\label{eq:relevant}
\eeq
apart from a dimensionless constant that is independent of $g$ and $T$. 
To check this, we plot $\ln(\tau_{\mathrm{adia}})$ as a function of $\ln(\Delta_{\mathrm{relevant}})$ in Fig.~\ref{fig:1Qubit-relevant-b}, which confirms the scaling predicted by Eq.~\eqref{eq:relevant} in the low $T$ regime.

More careful examination of Fig.~\ref{fig:1Qubit-relevant-a} reveals additional intriguing behavior, namely, the kink at $\ln(\Delta_{\mathrm{adia}})\approx 0$ for the high temperature ($T=0.5$) case. This kink is due to an eigenvalue crossing. To see this, first note that $\Delta_{\mathrm{adia}}$ depends on $g$ through $\Gamma \propto g^2$. Next, note that as $g$ becomes smaller, $|\lambda_{1,1}|=2|\Gamma|$ must become smaller than $|\lambda_{1,0}|=|\Gamma+i\delta|$, so that $\Delta_{\mathrm{adia}}$ [per Eq.~\eqref{eq:delta_adia_1q}] switches from the latter to the former, as illustrated in Fig.~\ref{fig:1Qubit-as-func-of-g}, resulting in a kink in $\tau_{\mathrm{adia}}$.

\begin{figure}[t] 
   \centering
\subfigure[]{\includegraphics[width=0.45\columnwidth]{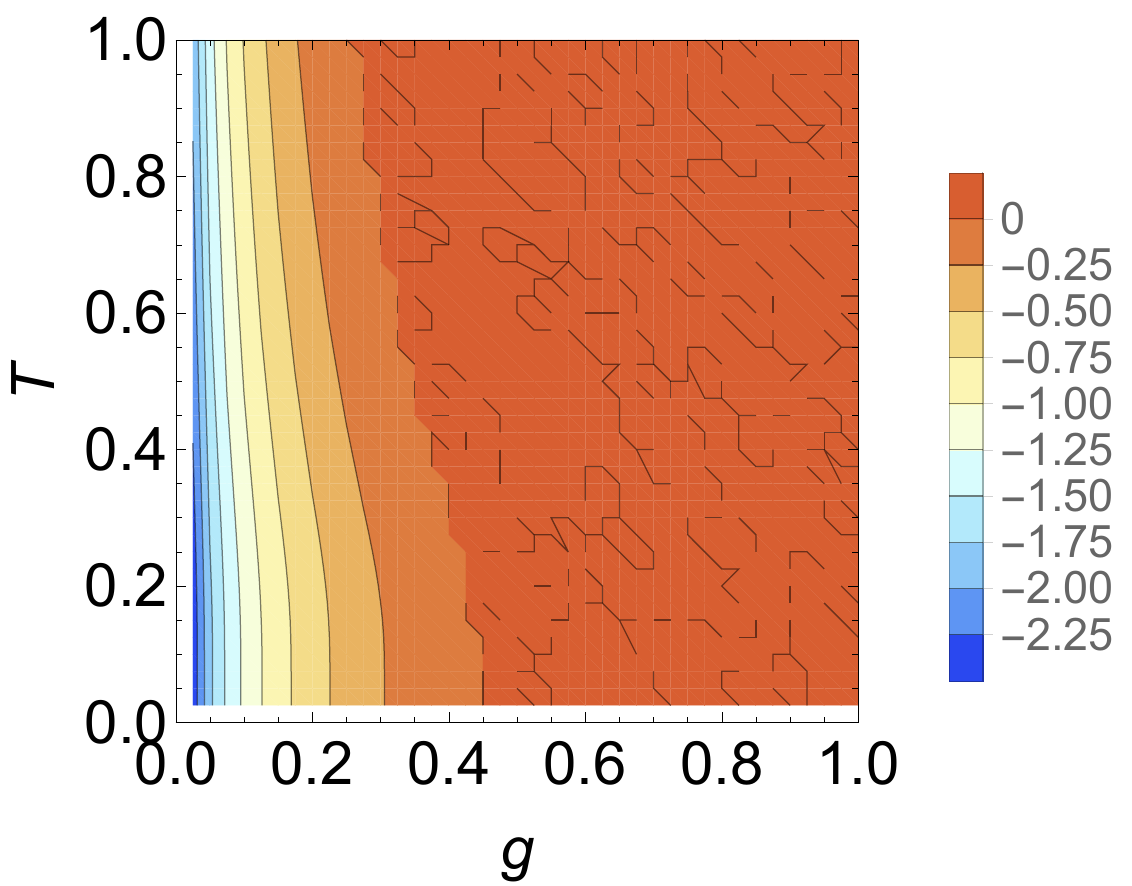}\label{fig:1Qubit-ratios-to-relevant-a}}
\subfigure[]{\includegraphics[width=0.45\columnwidth]{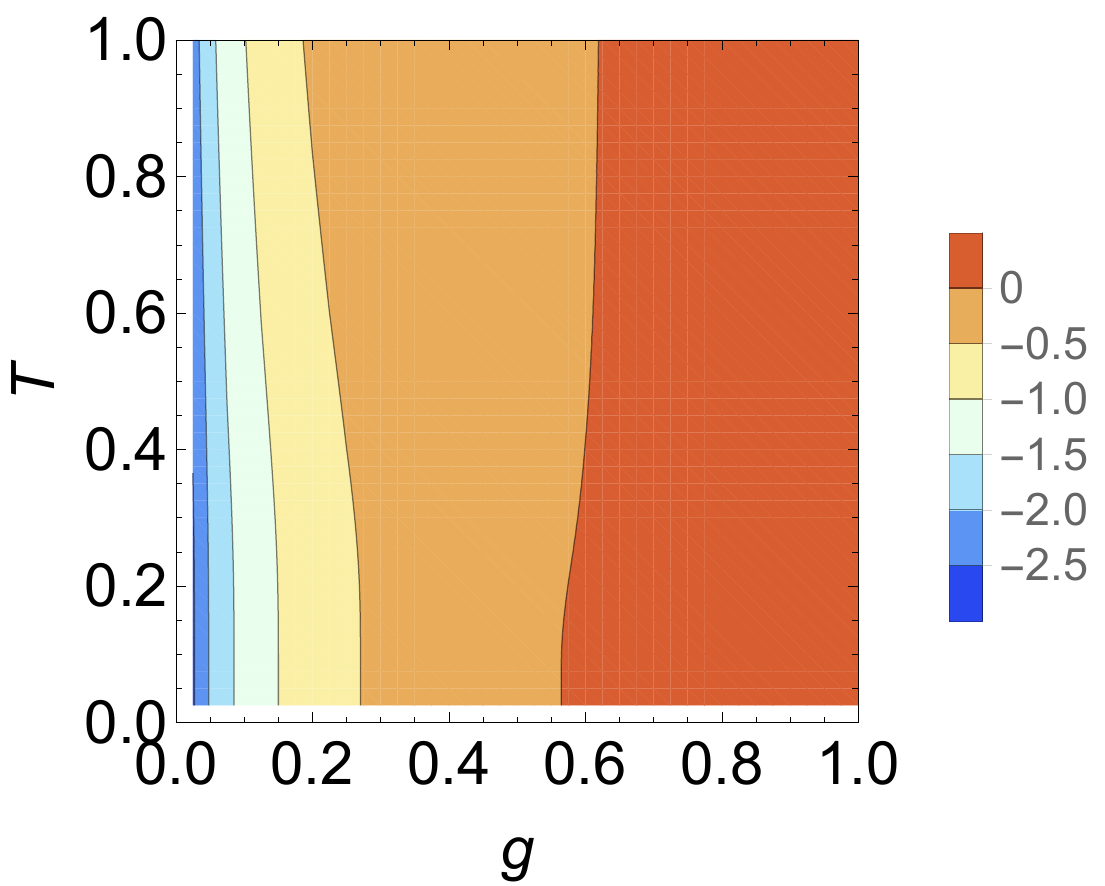}\label{fig:1Qubit-ratios-to-relevant-b}} 
   \caption{(Color online) (a) The $\log_{10}$ of the ratio $\Delta_{\mathrm{adia}}/\Delta_{\mathrm{relevant}}$. 
   (b) The $\log_{10}$ of the ratio $\Delta_{\mathrm{relax}}/\Delta_{\mathrm{relevant}}$.
The Hamiltonian minimum gap $\delta_{\min} = 1$. Both $T$ and $g$ are measured in units of $\delta_{\min}$.}
\label{fig:1Qubit-ratios-to-relevant}
\end{figure}

One may wonder how  $\Delta_{\mathrm{adia}}$ and $\Delta_{\mathrm{relevant}}$ compare in 
the
$(g,T)$ plane. Their ratio is shown in Fig.~\ref{fig:1Qubit-ratios-to-relevant-a}, and they can be seen to agree in the high $g$ and $T$ region. 
The agreement for high $g$ can be understood from the fact that solving for $| \lambda_1 | = | \lambda_{1,0} |$ at $T=0$, where $\Delta_{\mathrm{adia}} = \Delta_{\mathrm{relevant}}$, gives $g=1/(3^{1/4} \sqrt{\pi}) \approx 
0.43$. Figure~\ref{fig:1Qubit-ratios-to-relevant-a} also shows that $\Delta_{\mathrm{adia}}<\Delta_{\mathrm{relevant}}$ for $g<0.43$ at all temperatures, which is also where Fig.~\ref{fig:1Qubit-a} shows that adiabatic preparation is faster than relaxation, which means that $\Delta_{\mathrm{relevant}}$ is, indeed, the relevant gap in the weak coupling case. For comparison we plot $\Delta_{\mathrm{relax}}/\Delta_{\mathrm{relevant}}$ in Fig.~\ref{fig:1Qubit-ratios-to-relevant-b}, which more closely approximates $\tau_{\mathrm{adia}}/\tau_{\mathrm{relax}}$ [Fig.~\ref{fig:1Qubit-a}] than does the $\Delta_{\mathrm{relax}}/\Delta_{\mathrm{adia}}$ ratio [Fig.~\ref{fig:1Qubit-b}].

To conclude, as can be seen from Fig~\ref{fig:1Qubit}, for this model of a single qubit coupled to a thermal bath we find that at low temperatures and small coupling to the bath, the adiabatic process is more efficient than relaxation at preparing the steady state of the corresponding Lindbladian. Relaxation becomes more efficient than adiabatic preparation in the strong system-bath coupling regime. The differences between the two procedures gradually disappear as the temperature increases. For temperatures well below the minimum Hamiltonian gap, the adiabatic preparation time is captured well by $\Delta_{\mathrm{relevant}}$, but not by $\Delta_{\mathrm{adia}}$.

\subsection{The spike problem with a thermal bath} 
\label{sec:6}
%
We now consider the \qt{spike} problem, introduced in the 
closed system, 
$T=0$ setting in Ref.~\cite{Farhi-spike-problem}. This problem was designed to take classical single spin-flip simulated annealing exponentially longer to solve than adiabatic preparation of the ground state.  Here we generalize the problem to the $T>0$ setting, with the goal of preparing the thermal Gibbs state at time $t  = \tau$. We shall show that adiabatic preparation 
assisted by intermediate relaxation
can be significantly more efficient than 
pure
relaxation.  

The $n$-qubit Hamiltonian is the following:
\beq 
\label{eqt:spike}
H(t) = \left(1-\frac{t}{\tau} \right) \frac{1}{2}\sum_{i=1}^n \left( \1 - \sigma_i^x \right) +\frac{t}{\tau} \sum_{z \in \left\{0,1 \right\}^n} f(z) \left|{z} \rangle \langle{z} \right| \ ,
\eeq
where the cost function $f(z)$ is given by
\beq
f(z) = \left\{ \begin{array}{lr}
n \ ,& |z| = n/4 \\
|z| \ , & \mathrm{otherwise}
\end{array} \right. \ ,
\eeq
and $|z|$ denotes the Hamming weight of the classical bit-string $z$. The Hamiltonian is invariant under any permutation of the qubits. In order to  preserve this symmetry and keep simulations tractable, we consider a Davies-type open-system extension with the same symmetry. In particular, we choose a system-bath operator given by $S^y = \frac{1}{2} \sum_{i=1}^n \sigma_i^y$, and again use the generator of Eq.~\eqref{eq:Lind}, with an Ohmic spectral density as in Eq.~\eqref{eq:gamma}.

With these requirements, the instantaneous steady state in the $(n+1)$-dimensional totally symmetric (total spin $J = n/2$) subspace is given by:
\beq 
\label{eqt:rhoSS_Spike}
\rho_{\mathrm{SS}}(t) = \frac{1}{Z} e^{- \beta \left[H(t) \right]_{J = n/2}} \ ,
\eeq
where $\left[H(t) \right]_{J = n/2}$ is the Hamiltonian in Eq.~\eqref{eqt:spike} restricted to the symmetric subspace, and $Z = \mathrm{Tr}(e^{- \beta \left[H(t) \right]_{J = n/2}})$. 

For the adiabatic preparation, we choose the initial state to be $\rho_{\mathrm{SS}}(0)$. Since the initial state preserves the symmetry, the Lindbladian evolution does not take the system out of the symmetric subspace. 
For the relaxation-based preparation, we take the initial state to be the maximally mixed state in the symmetric subspace, so that the dynamics are again contained in an $(n+1)$-dimensional sector. 

\subsubsection{Closed-system results}

\begin{figure}[t] 
   \centering
   \subfigure[]{\includegraphics[width=0.45\columnwidth]{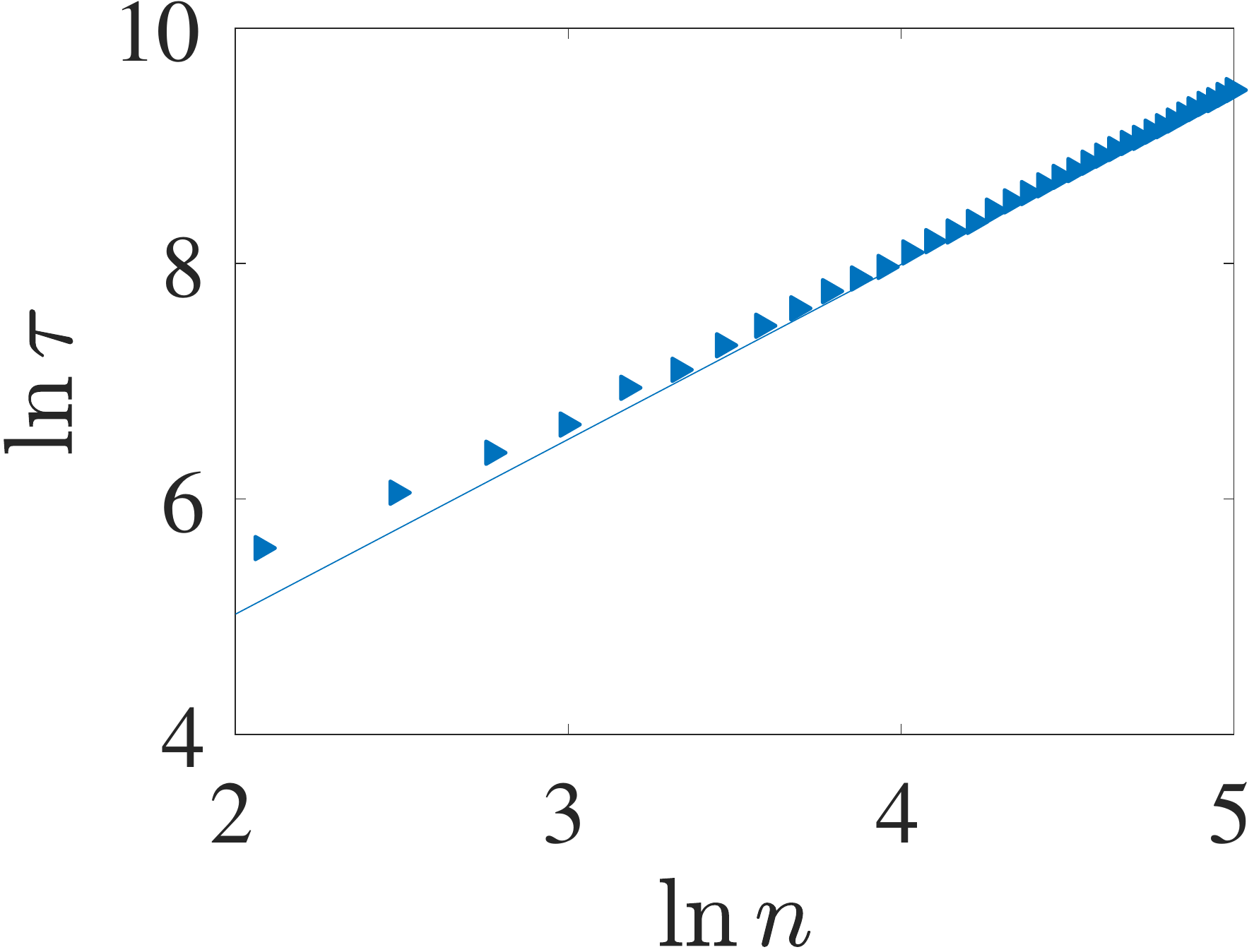}\label{fig:SpikeClosedSystem-a}}
   \subfigure[]{\includegraphics[width=0.45\columnwidth] {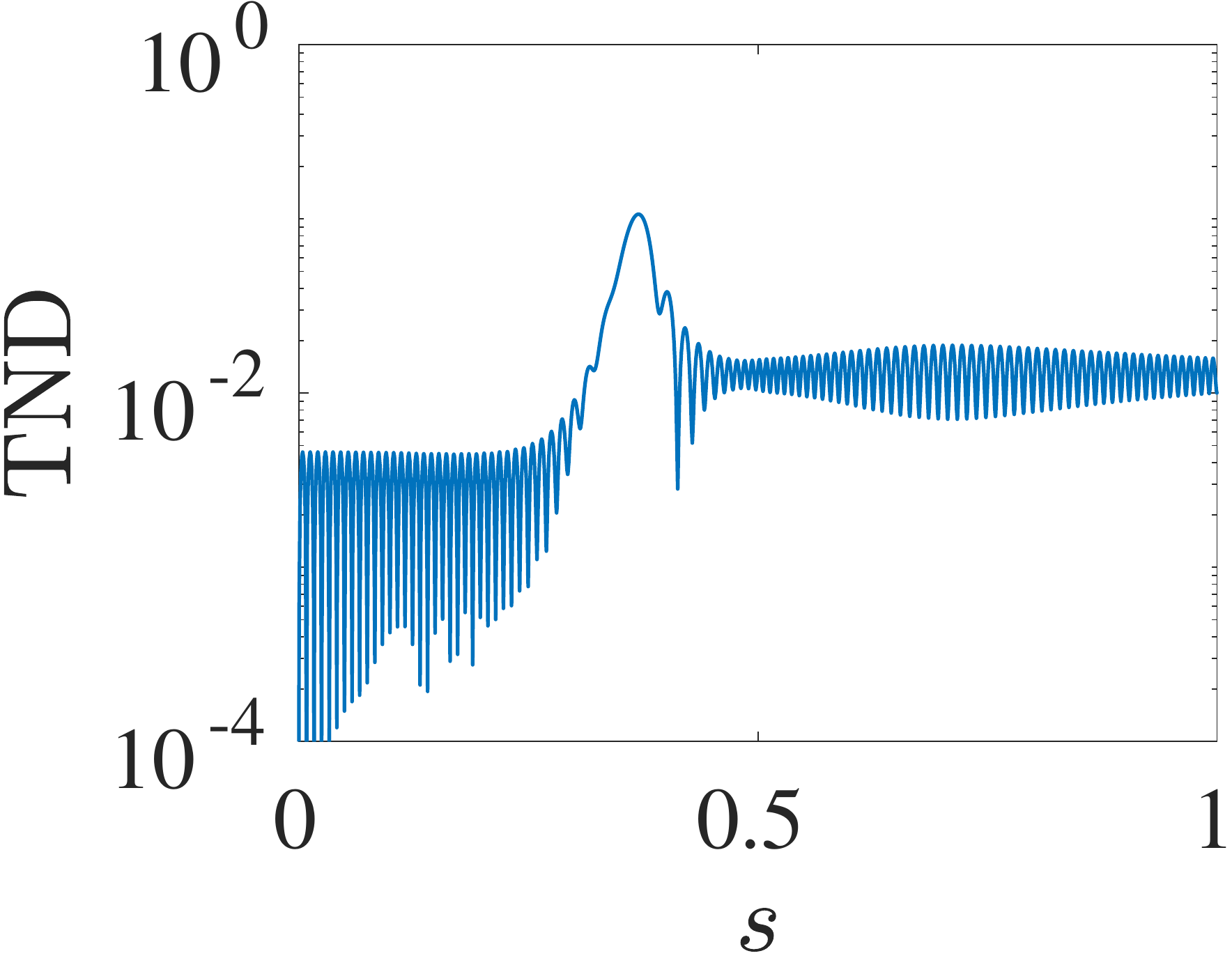}\label{fig:SpikeClosedSystem-b}}
   \caption{Closed-system simulation results for the spike
     problem. (a) The time required to reach a TND of $\epsilon =
     0.01$ from the ground state (the zero temperature Gibbs state)
     via evolution generated by the Hamiltonian~\eqref{eqt:spike}.
     The solid line is the best fit of $y=1.485x + 2.051$.  (b) The
     TND from the instantaneous zero-temperature Gibbs state as a
     function of the dimensionless parameter $s$ for $n = 20$ and
     $\tau = 759.5$.  The system remains very close to the
     instantaneous ground state ($> 0.99$ overlap squared) during the
     evolution. Here $\tau$ is measured in terms of the inverse of the minimum gap of $f(z)$.}  
   \label{fig:SpikeClosedSystem}
\end{figure}

We first provide in Fig.~\ref{fig:SpikeClosedSystem} simulation results for the scaling of the TND for the spike problem in the closed-system case with adiabatic preparation, where the criterion is to reach the final ground state (as opposed to the steady-state) with a fixed, high probability.  As seen in Fig.~\ref{fig:SpikeClosedSystem-a}, the time required to reach a given TND from the ground state scales polynomially. Moreover, the system remains close to the instantaneous ground state, for the parameters chosen in Fig.~\ref{fig:SpikeClosedSystem-b}. With these closed-system results in mind, let us now consider the open-system case, which exhibits strikingly different behavior.

\subsubsection{Open-system results}


\begin{figure}[t] 
\subfigure[]{\includegraphics[width=0.45\columnwidth]{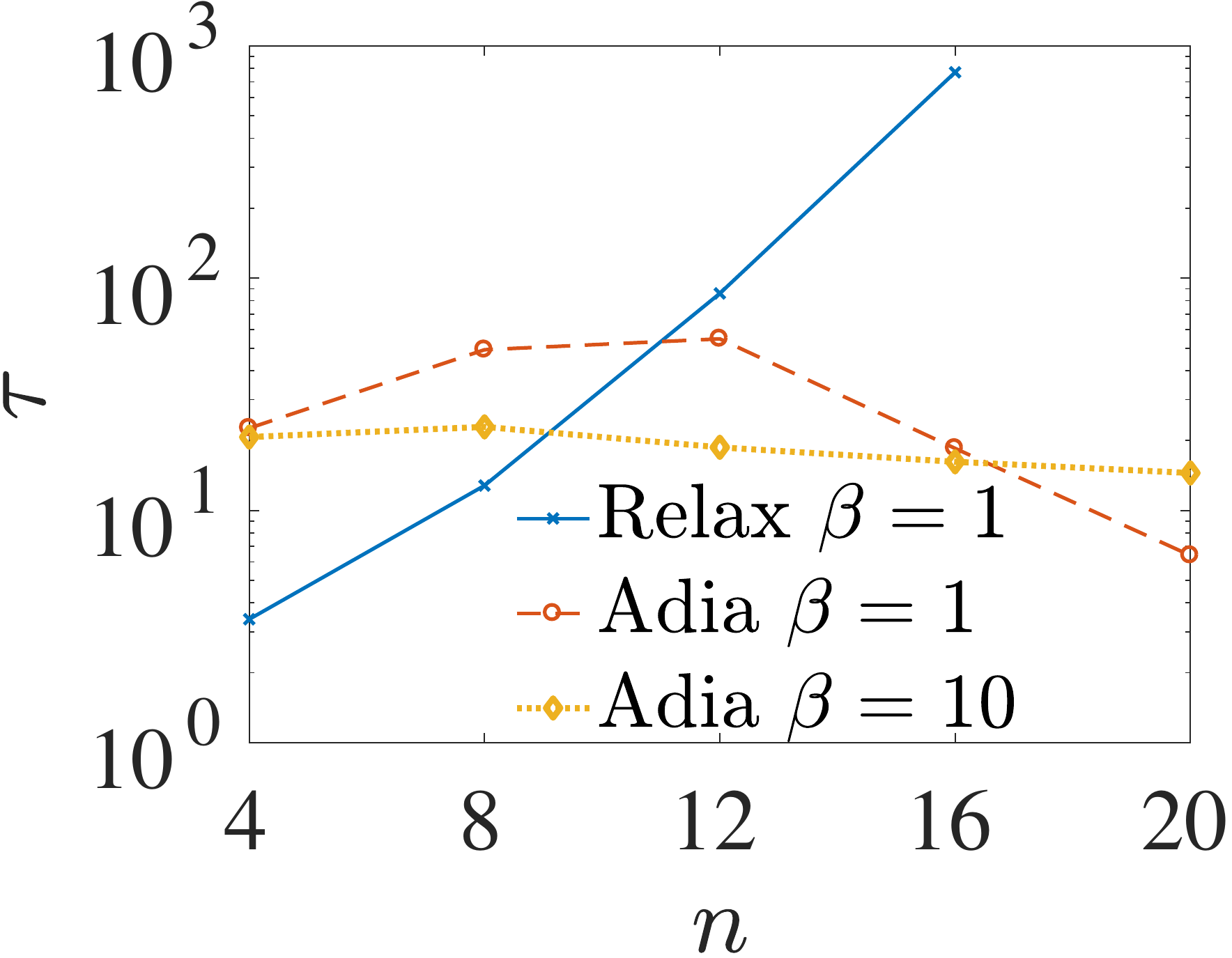} \label{fig:spike1}}
\subfigure[]{\includegraphics[width=0.46\columnwidth]{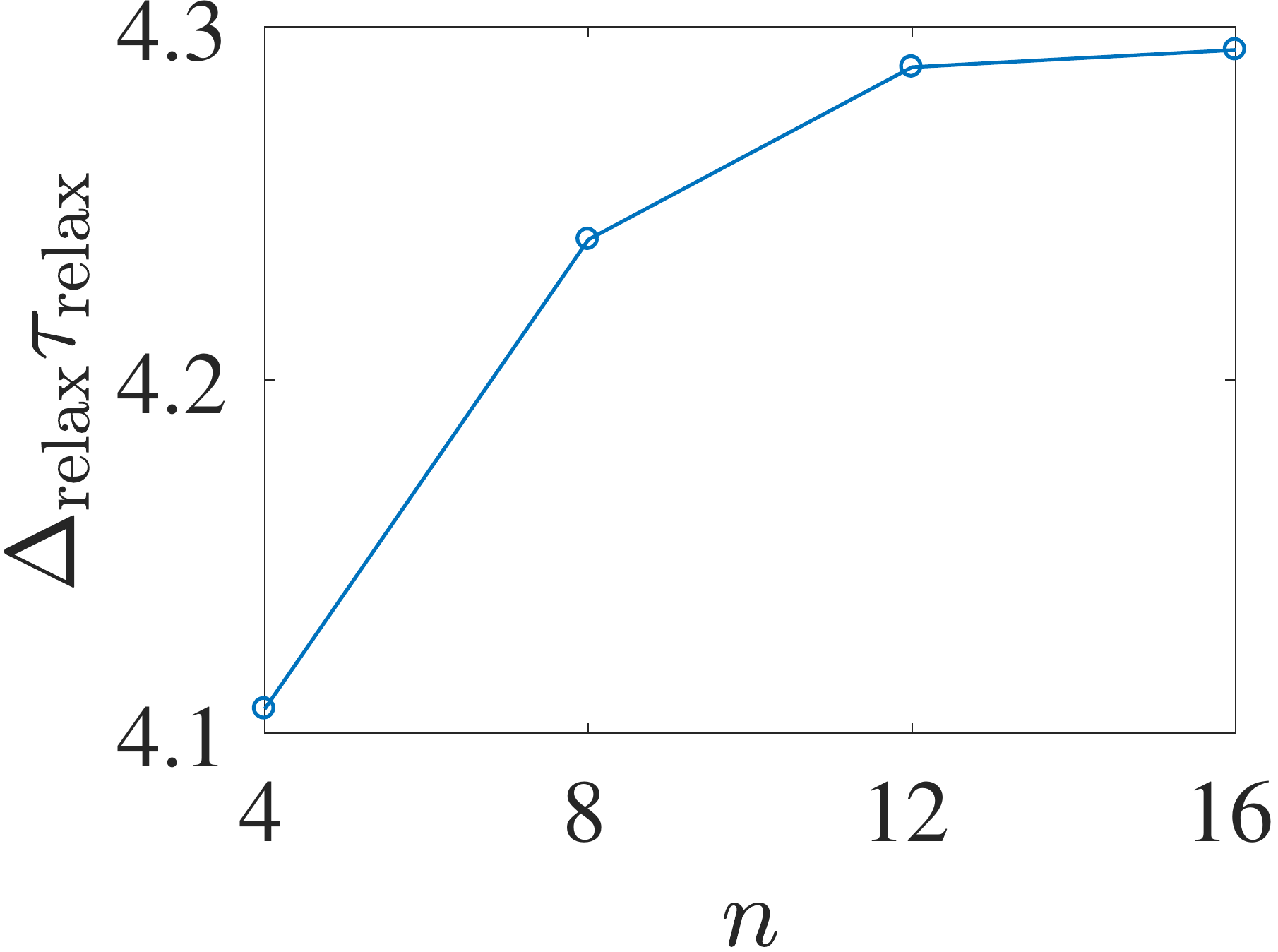} \label{fig:spike5}}
\caption{(Color online) (a) Time to reach a TND of $\epsilon = 10^{-2}$ for adiabatic preparation and relaxation, for spectral density [Eq.~\eqref{eq:gamma}] parameters $g=1$ and $\beta = 1, 10$. The relaxation time grows exponentially over the range of sizes shown, while the adiabatic preparation time \emph{decreases} with growing system size. (b) The product of the relaxation gap and the relaxation time for $\beta = 1, g=1$ exhibits polynomial scaling. The quantities $\tau$ and $\beta$ are measured in terms of the inverse of the minimum gap of $f(z)$, while $g$ is measured in terms of the minimum gap of $f(z)$.}
\label{fig:spike}
\end{figure}

\begin{figure}[t] 
   \subfigure[]{\includegraphics[width=0.45\columnwidth]{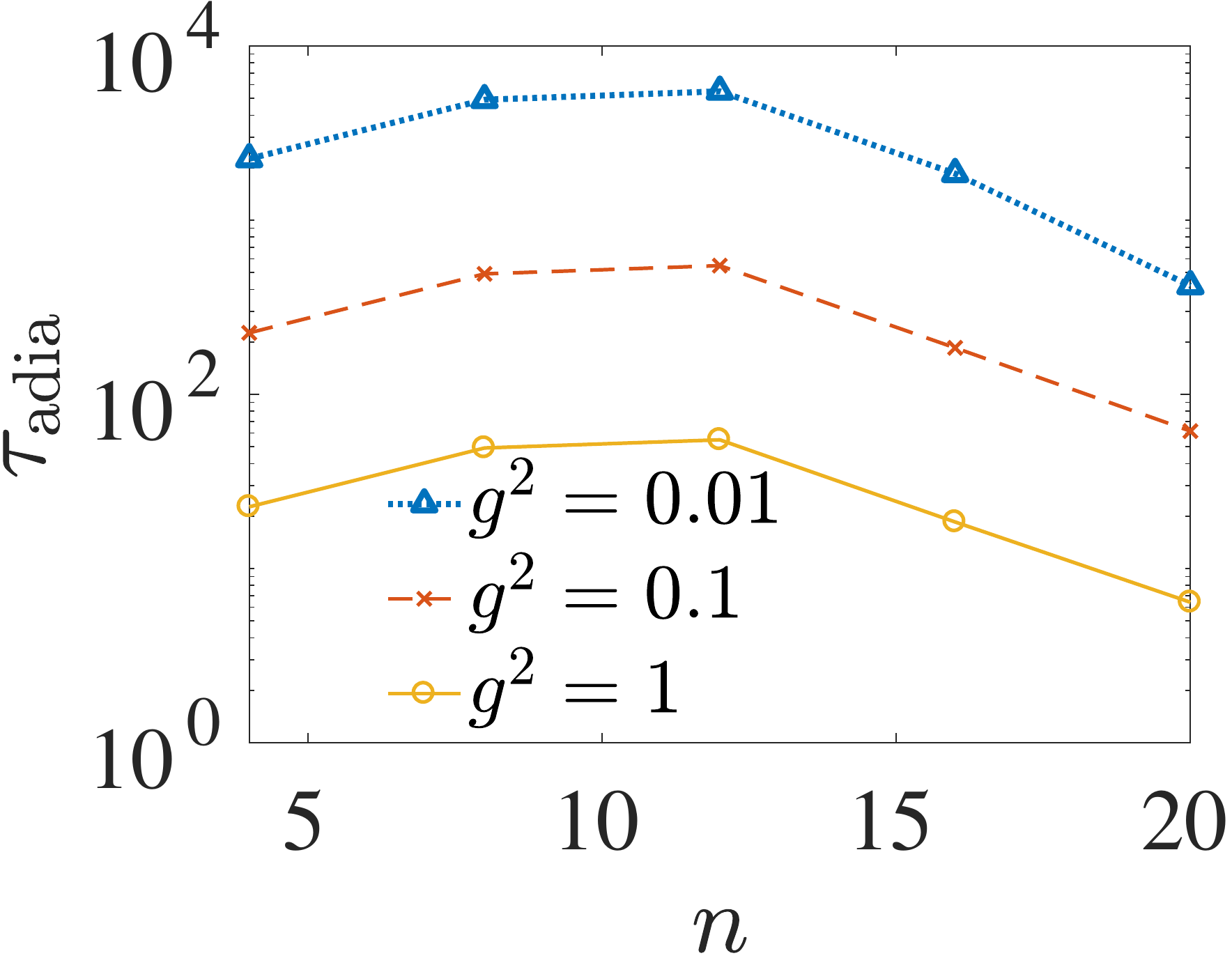} \label{fig:adiaSpike_beta=1}}
    \subfigure[]{\includegraphics[width=0.45\columnwidth]{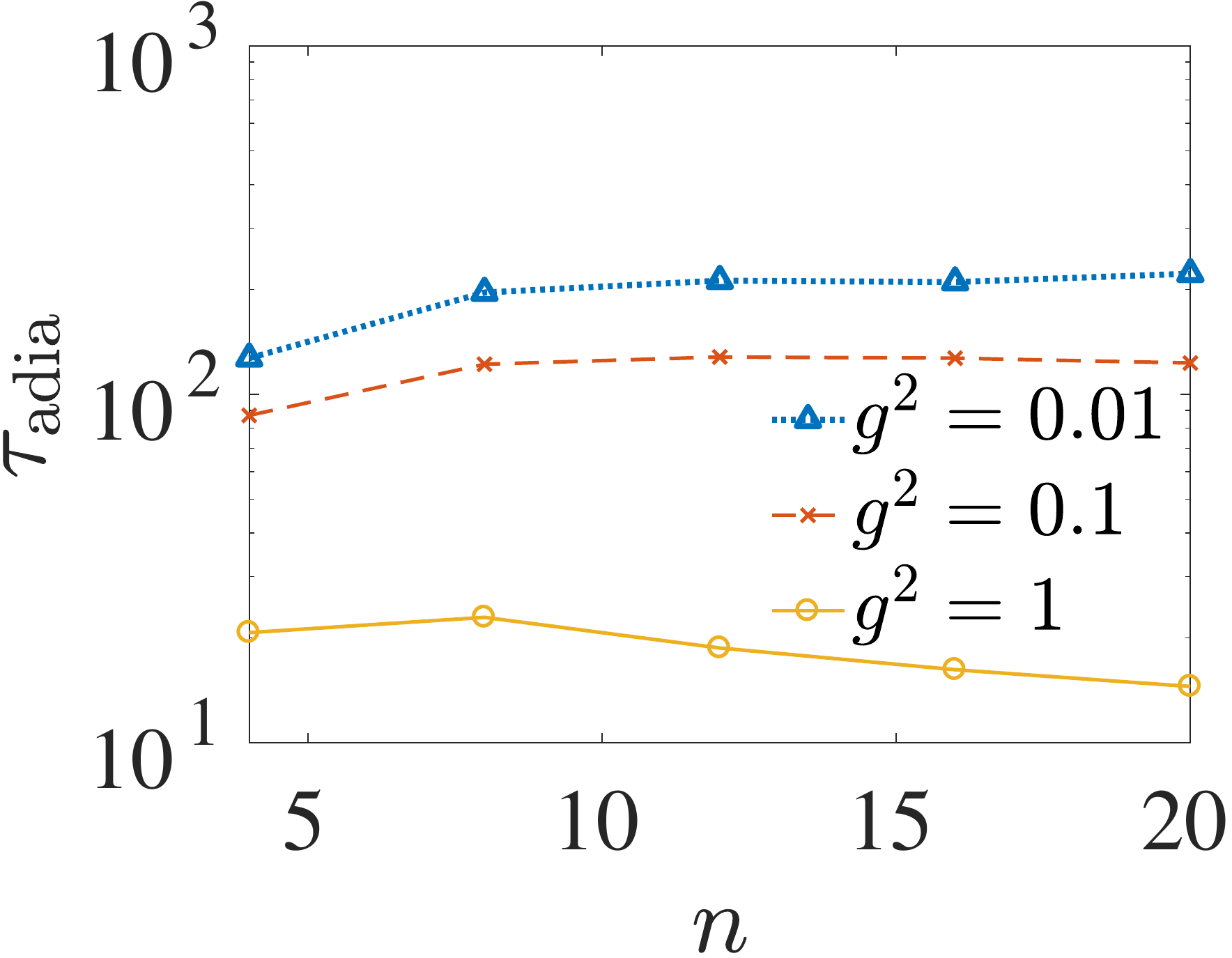} \label{fig:negativeScaling} }
\subfigure[]{\includegraphics[width=0.45\columnwidth]{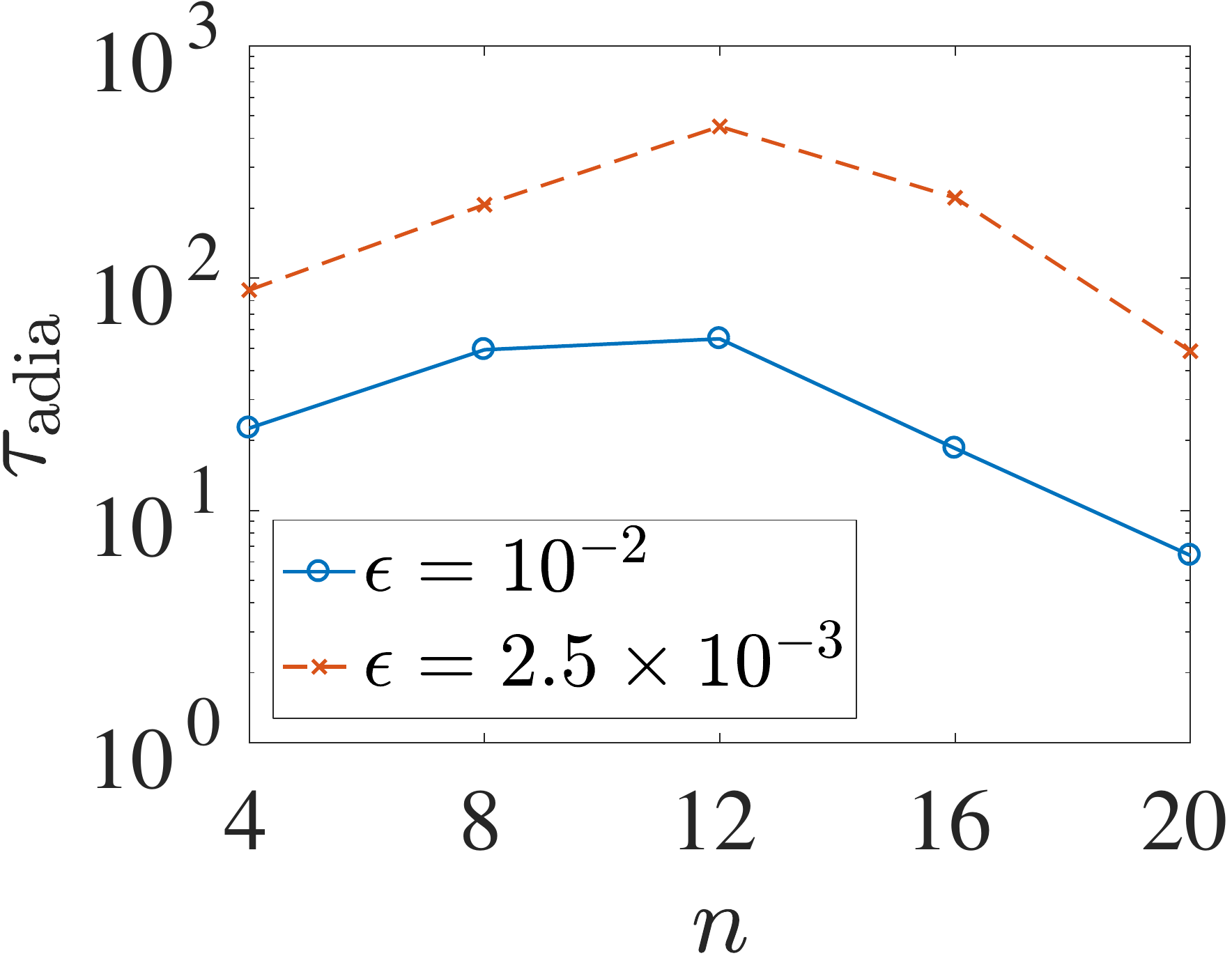}    \label{fig:spikeExtra3}}
   \subfigure[]{\includegraphics[width=0.45\columnwidth]{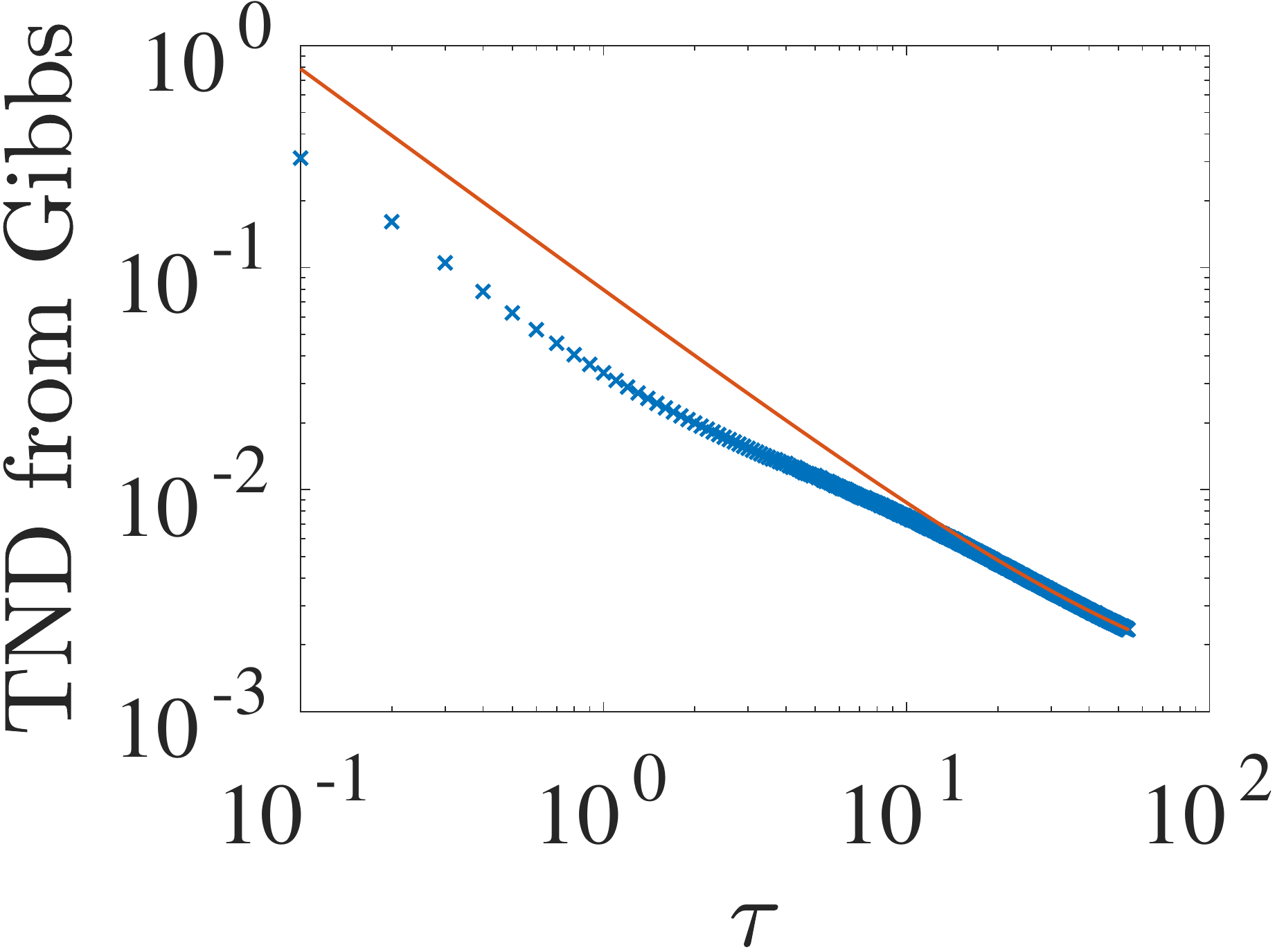}    \label{fig:spikeExtra6}}
   \caption{(Color online) (a)-(c) Time to reach a given TND of $\epsilon$ for adiabatic preparation of the \qt{spike} thermal state.  (a) $\beta = 1$ and $\epsilon = 10^{-2}$, (b) $\beta = 10$ and $\epsilon = 10^{-2}$, (c) $\beta = 1$ and $g=1$. (d) TND between the evolved state and the final thermal state for $n=20$ and $g=\beta = 1$.  The solid line is a best fit to $a + b /\tau$.  Same units as in Fig.~\ref{fig:spike}.}
   \label{fig:spikeExtra}
\end{figure}
%


We show results comparing the TTSS for relaxation and adiabatic preparation in Fig.~\ref{fig:spike1}. For the relaxation process, we observe an exponential growth in $\tau_{\mathrm{relax}}$ over the range of sizes tested.
 In Fig.~\ref{fig:spike5} we confirm our prediction for the TTSS for the relaxation process [Eq.~\eqref{eq:T_relax}] in that the product $\Delta_{\mathrm{relax}} \tau_{\mathrm{relax}}$ is a polynomial in $n$.  

Counterintuitively, the adiabatic preparation process displays a \emph{negative} TTSS slope with the number of sites $n$, 
i.e., $\tau_{\mathrm{adia}}$ decreases with increasing $n$.   The negative slope is more dramatic
at higher temperatures ($\beta = 1$) but remains negative even at
lower temperatures ($\beta = 10$) and sufficiently high $g$. We check the dependence of the scaling of $\tau_{\mathrm{adia}}$ on both $g$ and $\beta$ in Fig.~\ref{fig:spikeExtra}. As discussed in more detail below [Fig.~\ref{fig:9}], for high temperatures the evolution can be seen to be very close to adiabatic in the open-system sense, so we do not expect the scaling behavior to change qualitatively as we continue to vary $g$. Indeed, Fig.~\ref{fig:adiaSpike_beta=1} shows that the negative scaling persists for high temperature ($\beta = 1$) for $g^2$ values spanning two orders of magnitude. The $\tau_{\mathrm{adia}}$ required for a given TND likewise increases by two orders of magnitude, which is consistent with the dissipative dynamics being dominant, since the strength of thermal transitions is governed by $g^2 \tau$.   

For sufficiently low temperature ($\beta=10$) and sufficiently small $g$ the scaling does becomes positive, as can be seen in Fig.~\ref{fig:negativeScaling}. This indicates that in this parameter range the unitary dynamics must still play an important role. Indeed, for $\beta = 10$, the thermal state has a high overlap of $0.92$ with the instantaneous ground state at the point where the Hamiltonian gap is minimized.

To ensure that the negative scaling result is not an artifact of the specific choice of TND precision, we study the dependence on $\epsilon$ in Fig.~\ref{fig:spikeExtra3}.  Decreasing $\epsilon$ increases the time required to reach the desired target, but 
since Fig.~\ref{fig:spikeExtra6} shows that the TND scales (for a fixed $n$) as $1/\tau$, this does not change the qualitative scaling of the TND as $n$ is varied, at least at high temperatures.

The striking aspect of our results is that, especially in the low temperature regime,
they seem to indicate a preparation time that improves upon the
closed-system case.  A key difference
between the two cases is that, for the same error $\epsilon$, 
in the adiabatic preparation of 
thermal states at positive temperature, the system deviates from the
instantaneous steady state by a larger amount compared to the
closed-system case [compare Fig.~\ref{fig:spikeExtra5b} with
  Fig.~\ref{fig:SpikeClosedSystem-b}].  
This is intuitively explained by there being additional (relaxation)
channels that increase the population of the steady state for $T>0$,
as compared to the closed-system case (subject to the caveat that our
simulations are limited to $n\leq 20$).

\subsubsection{Explanation of the open system results}

\begin{figure}[t] 
\subfigure[]{\includegraphics[width=0.48\columnwidth]{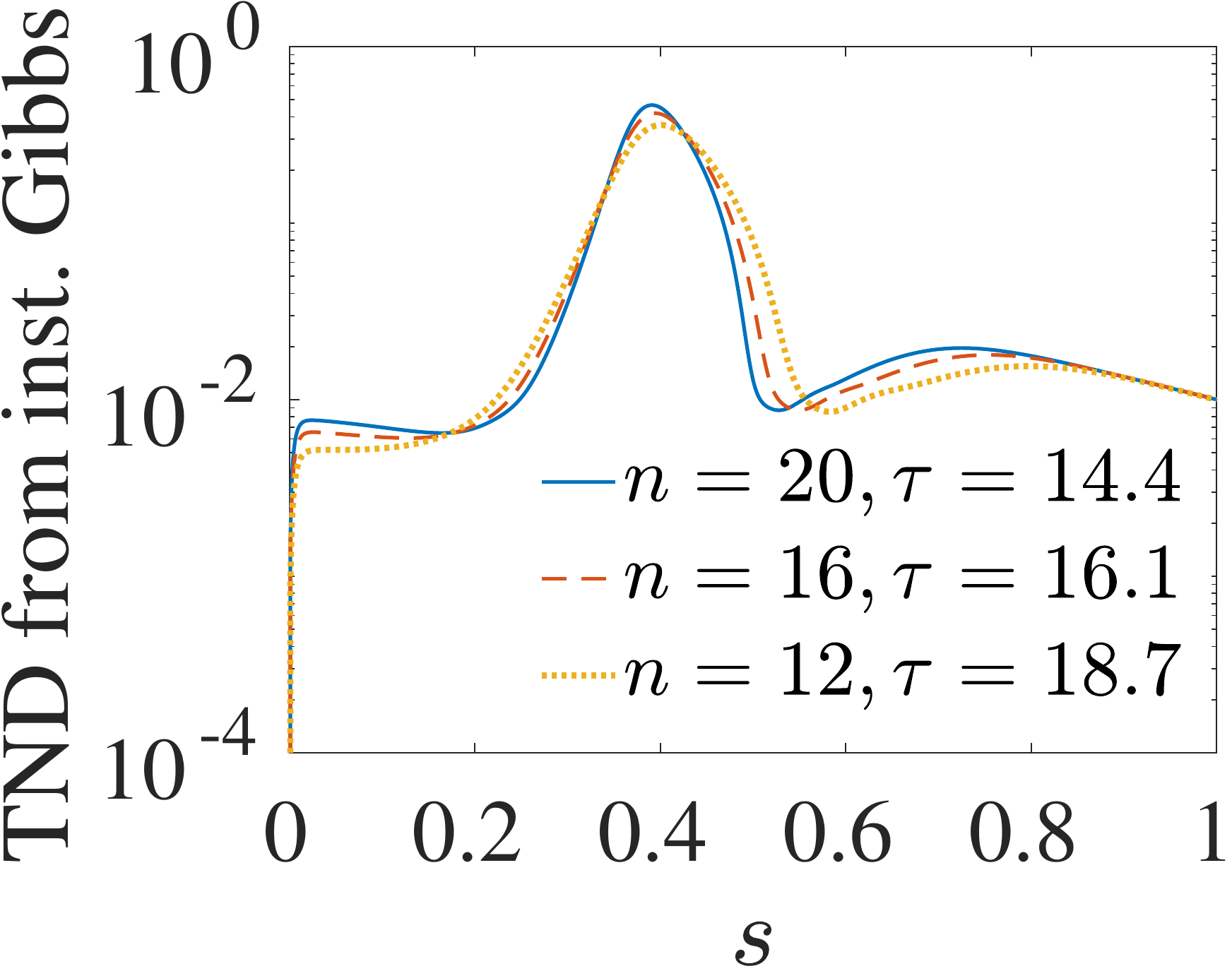} \label{fig:spikeExtra5b}} 
\subfigure[]{\includegraphics[width=0.48\columnwidth]{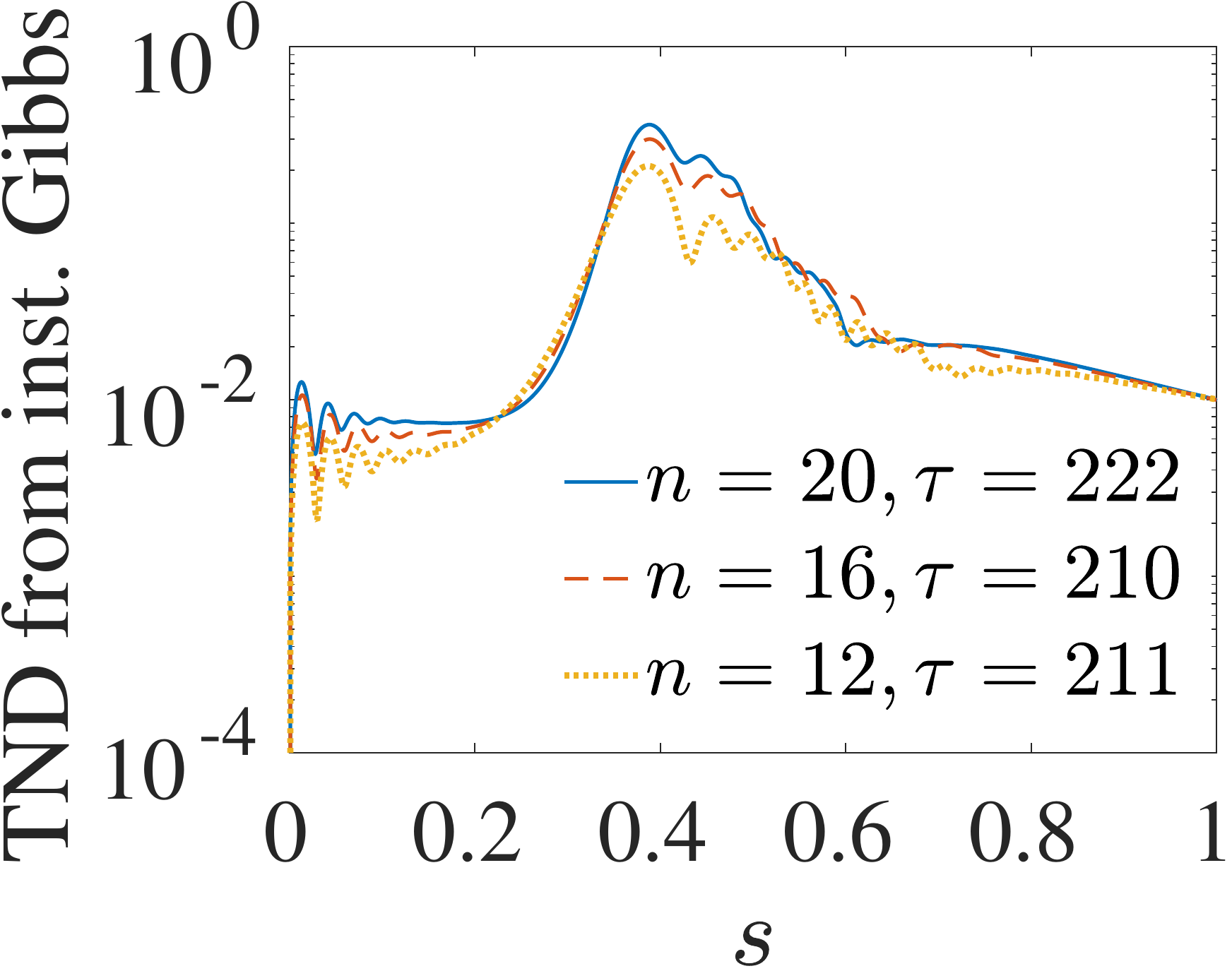} \label{fig:spikeExtra5a}} 
\subfigure[]{\includegraphics[width=0.48\columnwidth]{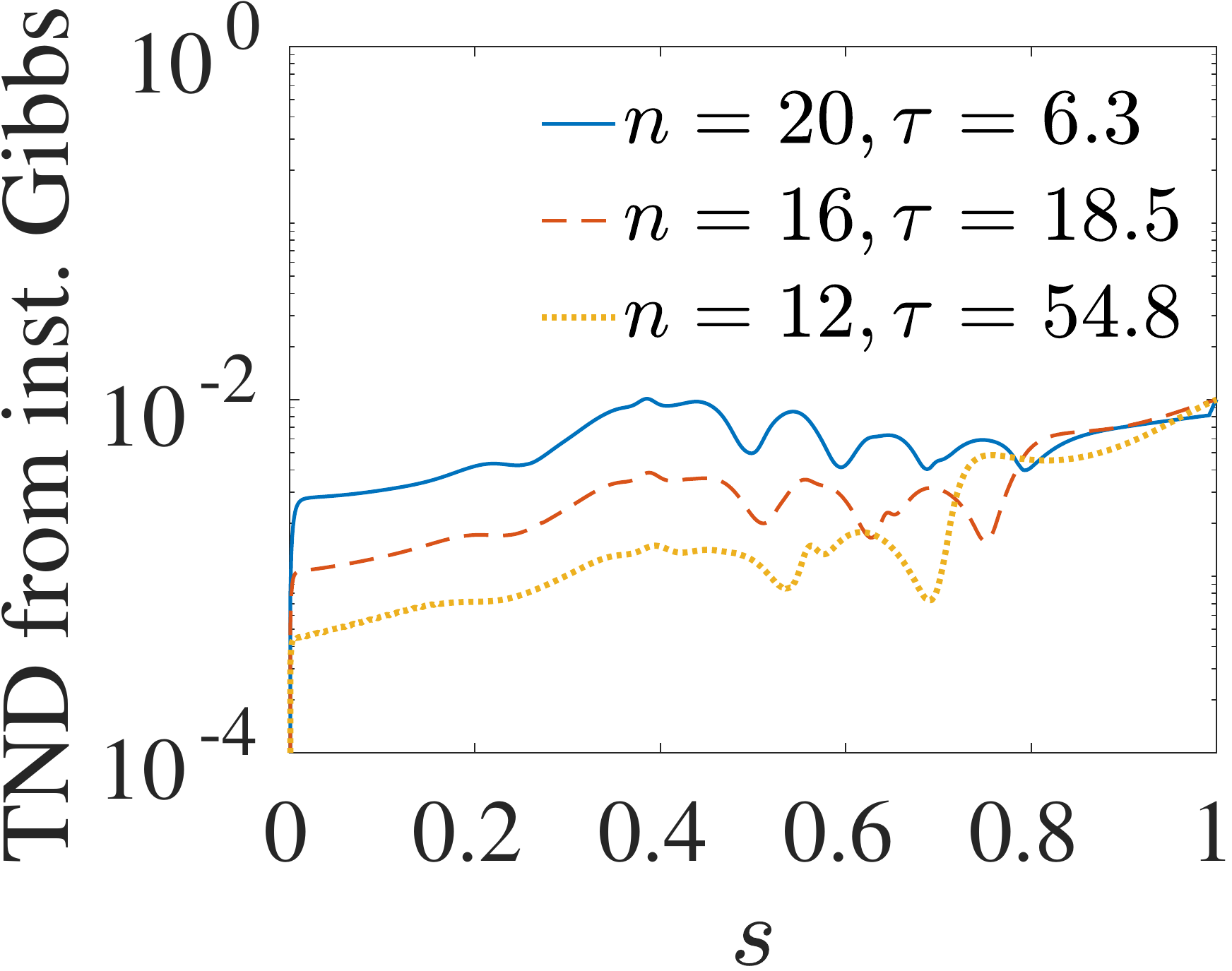} \label{fig:spikeExtra5c}} 
\subfigure[]{\includegraphics[width=0.48\columnwidth]{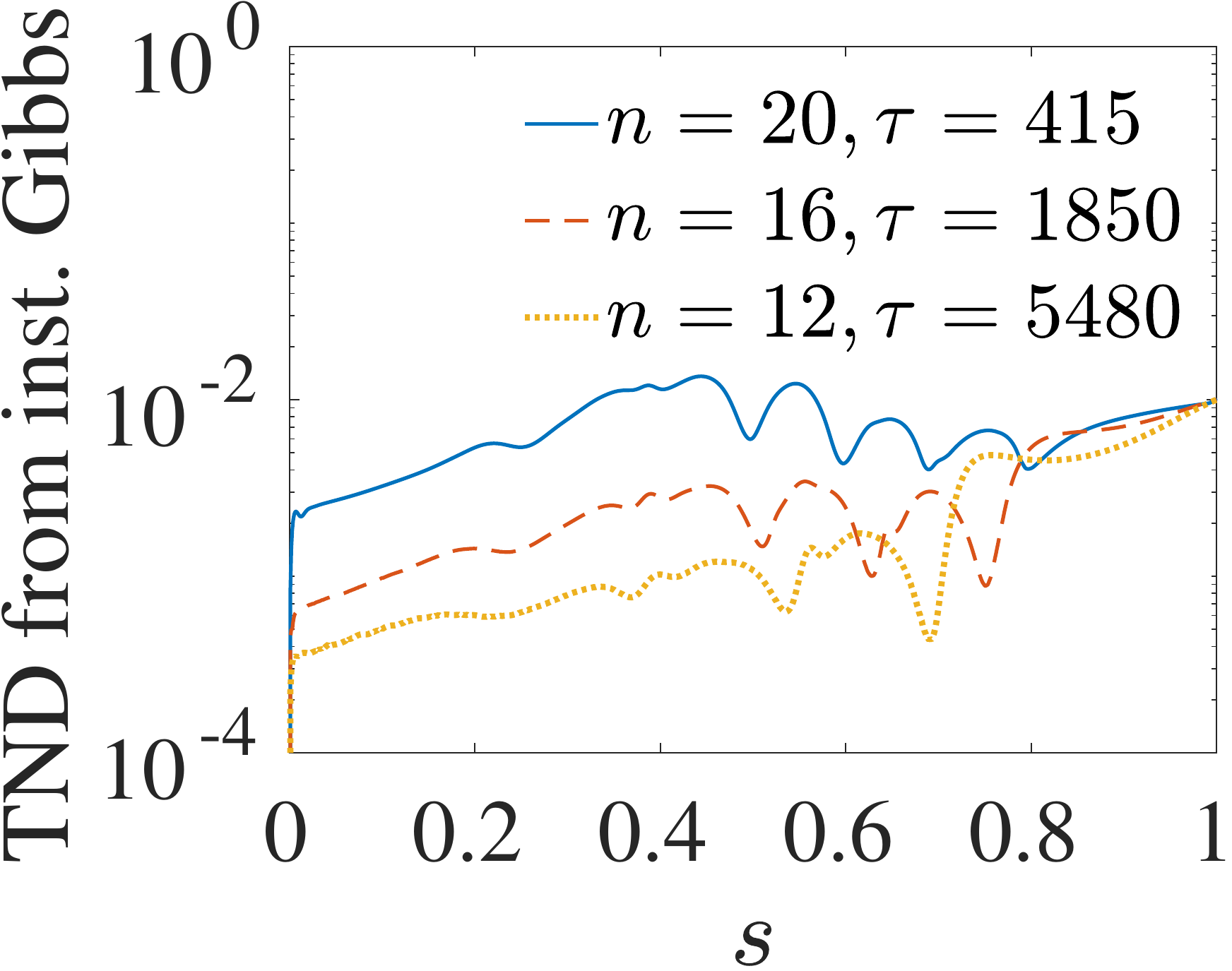} \label{fig:spikeExtra5d}} 
   \caption{(Color online) TND of the evolved state from the instantaneous thermal state for $\beta = 10$ (left) and $\beta=1$ (right) and system-bath coupling $g^2=1$ (top), and $g^2=10^{-2}$ (bottom).  For all $n$ values shown, $\tau$ is chosen so that $\epsilon = 10^{-2}$.  Same units as in Fig.~\ref{fig:spike}.}
\label{fig:9}
\end{figure}
\begin{figure}[t] 
   \subfigure[]{\includegraphics[width=0.49\columnwidth]{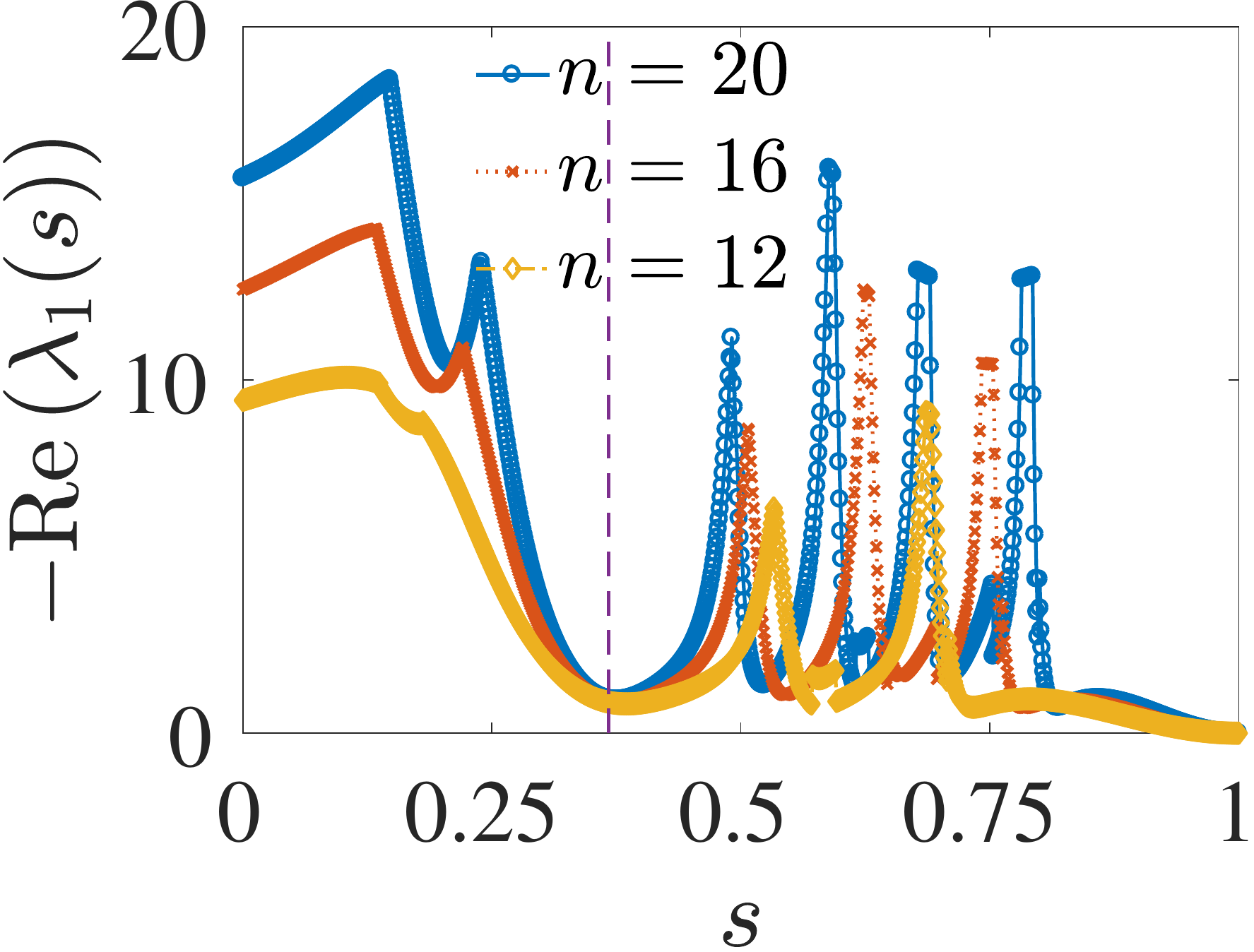} }
    \subfigure[]{\includegraphics[width=0.49\columnwidth]{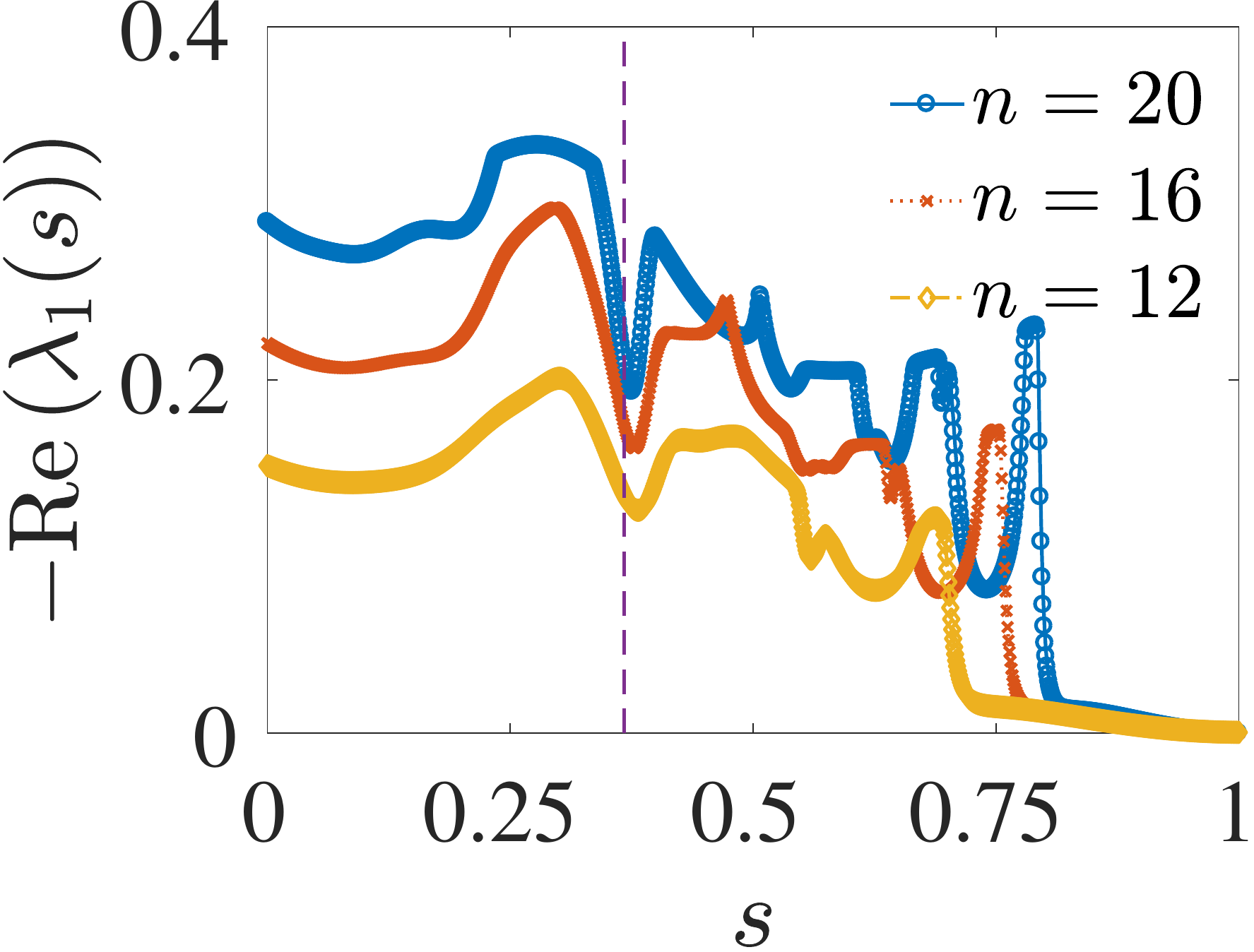}}\\
   \subfigure[]{\includegraphics[width=0.49\columnwidth]{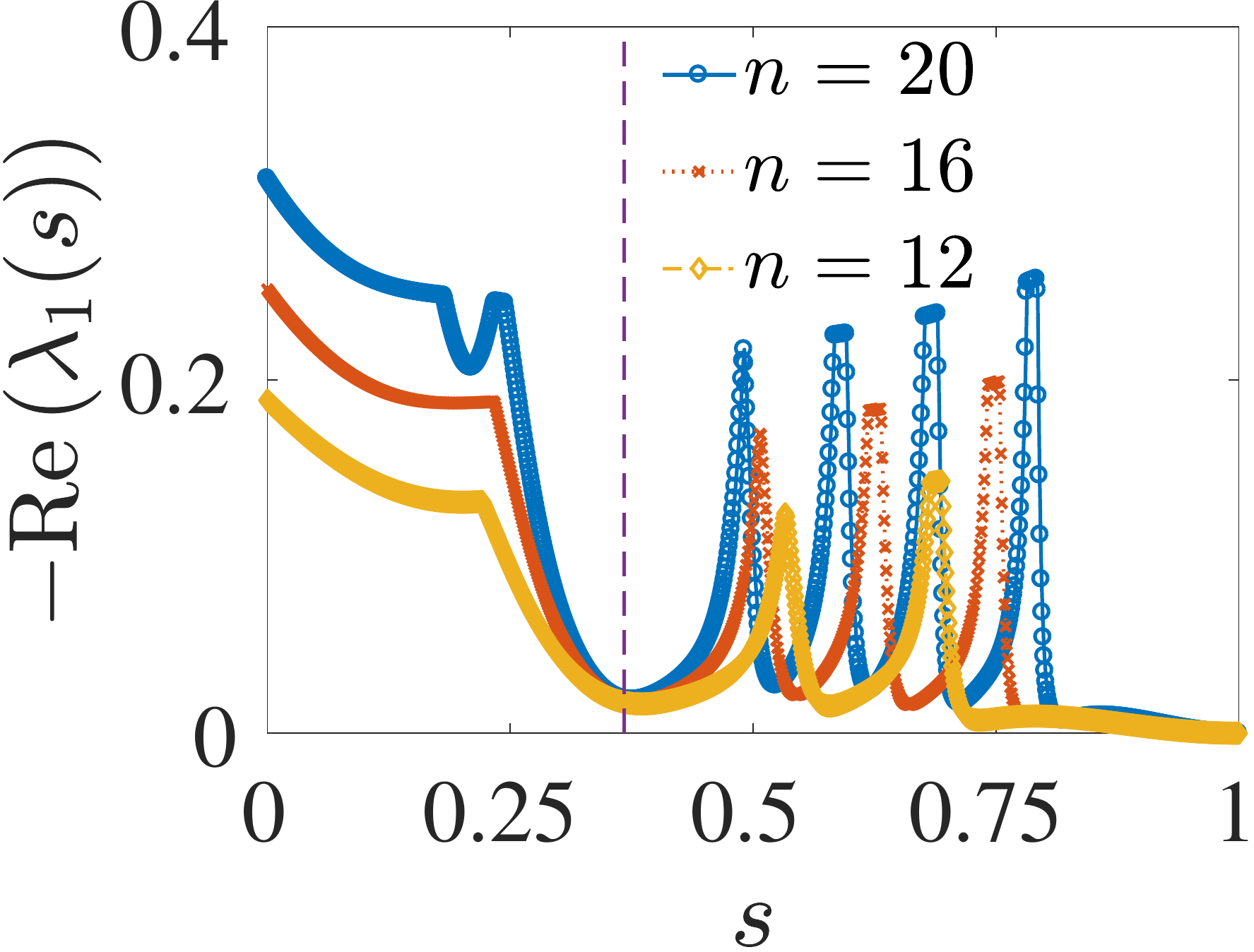} }
    \subfigure[]{\includegraphics[width=0.49\columnwidth]{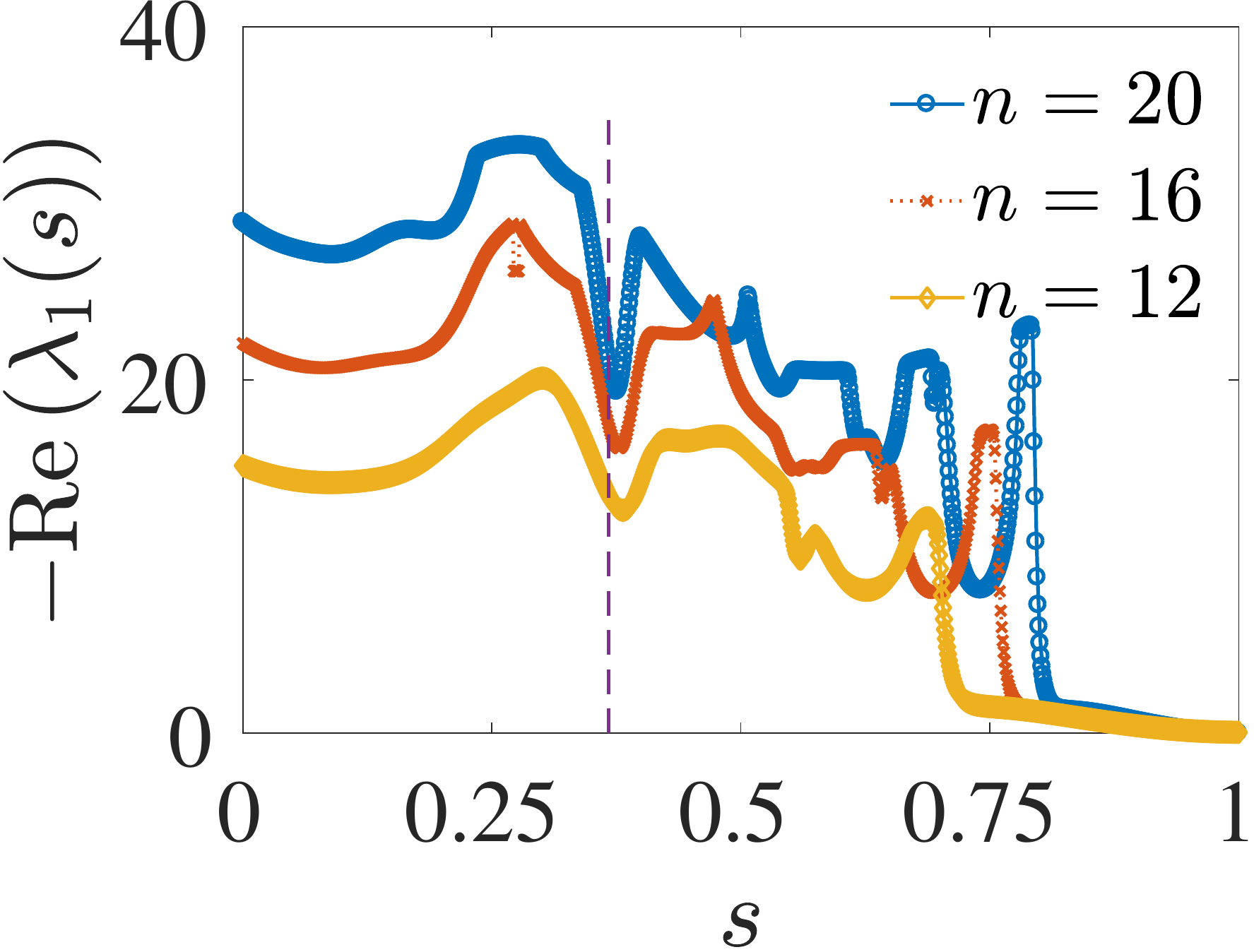}}
   \caption{(Color online) Real part of the Lindbladian gap as a function of the dimensionless time $s$ for $\beta = 10$ (left), and $\beta = 1$ (right). The system-bath coupling is $g^2=1$ (top), and $g^2=10^{-2}$ (bottom).  The vertical dashed line is the location of the minimum Hamiltonian gap.  The slope discontinuities are due to eigenvalue crossings. Same units as in Fig.~\ref{fig:spike}.}  
\label{fig:EigevaluesSpikeSize}
\end{figure}

In order to understand why the adiabatic preparation becomes more
efficient as the system size grows for sufficiently low $T$, we consider what happens during the adiabatic evolution, i.e., as a function of $s=t/\tau$.

In Fig.~\ref{fig:9} we plot the instantaneous TND between the evolving state and the instantaneous steady (thermal) state for three different problem sizes, and for different $(g,T)$ combinations. The TND hardly exceeds $10^{-2}$ for $\beta = 1$ (right column), indicating that the evolution is close to being adiabatic in the open system sense throughout the evolution.  For $\beta = 10$ (left column), the TND becomes large at intermediate points in the evolution but then drops to very small values. Thus, while the evolution is not adiabatic throughout, relaxation processes can quickly bring the state back to being close to the steady state, an effect that is more pronounced in the lower temperature case. This beneficial effect of \emph{relaxation throughout the adiabatic evolution} is a key operative mechanism (see also Refs.~\cite{TAQC,DWave-16q}) that helps to explain the advantage of adiabatic preparation over the relaxation strategy at $t=\tau$. Indeed, it is important to draw a clear distinction between relaxation-assisted adiabatic evolution and a purely relaxation-based strategy.

Studying the Lindbladian gap during the evolution clarifies that what is happening in Fig.~\ref{fig:9} is indeed a relaxation-assisted return to the instantaneous steady state. We present this analysis in Fig.~\ref{fig:EigevaluesSpikeSize}, which displays $\eta_1(s)=-\mathrm{Re}[\lambda_1(s)]$, where $\lambda_1(s)$ is the non-zero Lindbladian eigenvalue with the smallest modulus, for three different system sizes. As established in Sec.~\ref{sec:2}, this quantity determines the relaxation rate towards the instantaneous steady state.

It turns out that $\mathrm{Im}[\lambda_1(s)] = 0$ $\forall s\in[0,1]$ for the cases shown in Fig.~\ref{fig:EigevaluesSpikeSize}.\footnote{The reason is similar to the eigenvalue crossing phenomenon seen in the single-qubit case, shown in Fig.~\ref{fig:1Qubit-as-func-of-g-b}.} We first note that since, as seen in Fig.~\ref{fig:EigevaluesSpikeSize}, $-\mathrm{Re}[\lambda_1(s)]$ is minimized at $s=1$ ($t=\tau$), we find that for this
problem $\Delta_{\mathrm{adia}} = \Delta_{\mathrm{relax}}$. Coupled with the scaling seen in Fig.~\ref{fig:spike}, where adiabatic preparation bests relaxation, this confirms once more that the pessimistic prediction for the adiabatic preparation time obtained by contrasting Eqs.~\eqref{eq:T_relax} and \eqref{eq:T_adia}, cannot be correct.  

Apart from the vanishing of $\mathrm{Re}[\lambda_1(1)]$, Fig.~\ref{fig:EigevaluesSpikeSize} exhibits much additional structure, which we analyze next. Note that $-\mathrm{Re}[\lambda_1(s)]$ has a pronounced local minimum at the point $s^*$ where the Hamiltonian gap is minimized (the dashed lines). This point is also (roughly) where the TNDs in Fig.~\ref{fig:9} peak, which is sensible since $-\mathrm{Re}[\lambda_1(s)]$ determines the relaxation rate towards the instantaneous thermal state. Focusing on the $\beta=10$ case [Fig.~\ref{fig:EigevaluesSpikeSize}], we observe that a succession of large peaks for $s>s^*$ corresponds to strong relaxation events, which explain why the TND drops sharply for $s>s^*$ and $\beta=10$ in Fig.~\ref{fig:9}. Moreover, the peak height increases with the problem size $n$, which clearly signals that relaxation-assistance plays a more pronounced role as the problem size increases, and also explains why the relaxation-assisted adiabatic preparation becomes more efficient as $n$ grows. This, then, is the sought-after explanation for why adiabatic preparation exhibits a negative scaling with problem size, as seen in Fig.~\ref{fig:spike1}. 

The effect of relaxation-assistance is even more pronounced in the higher temperature case ($\beta=1$) seen in the right column of Fig.~\ref{fig:EigevaluesSpikeSize}, where $-\mathrm{Re}[\lambda_1(s)]$ is much larger for most $s$ values than in the low temperature case ($\beta=10$). Correspondingly, as seen in Fig.~\ref{fig:9}, the TND is significantly smaller for most $s$ values in the $\beta=1$ case than in the $\beta=10$ case, and at the same time size plays a much more significant role for $\beta=1$ than for $\beta=10$. This explains why the negative scaling with problem size [Fig.~\ref{fig:spike1}] is also more pronounced. 


\subsubsection{Instantaneous adiabaticity \textit{vs} final time adiabaticity}

\begin{figure}[t] %
   \subfigure[]{\includegraphics[width=0.49\columnwidth]{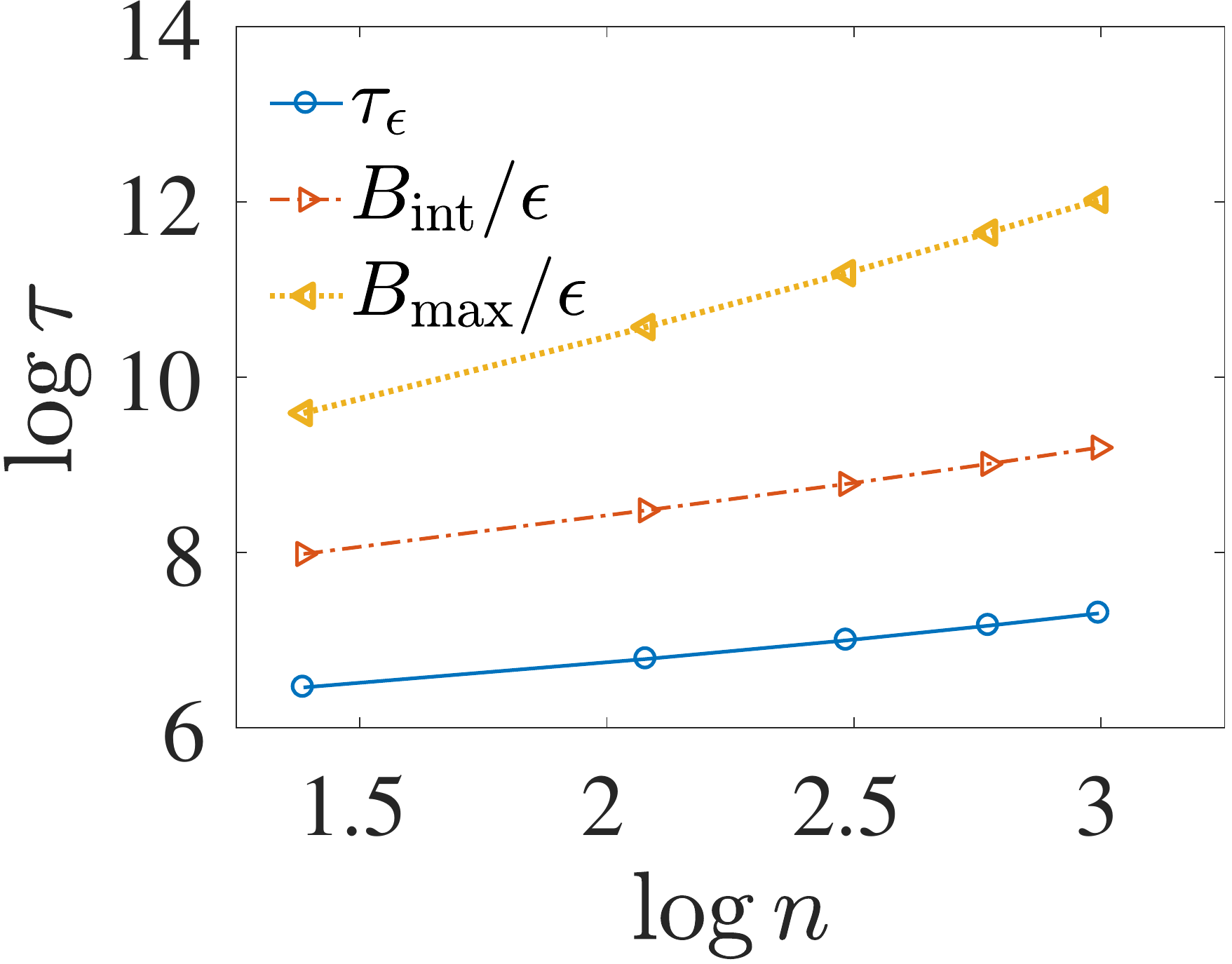}\label{fig:adiabaticSpike}}
   \subfigure[]{\includegraphics[width=0.49\columnwidth]{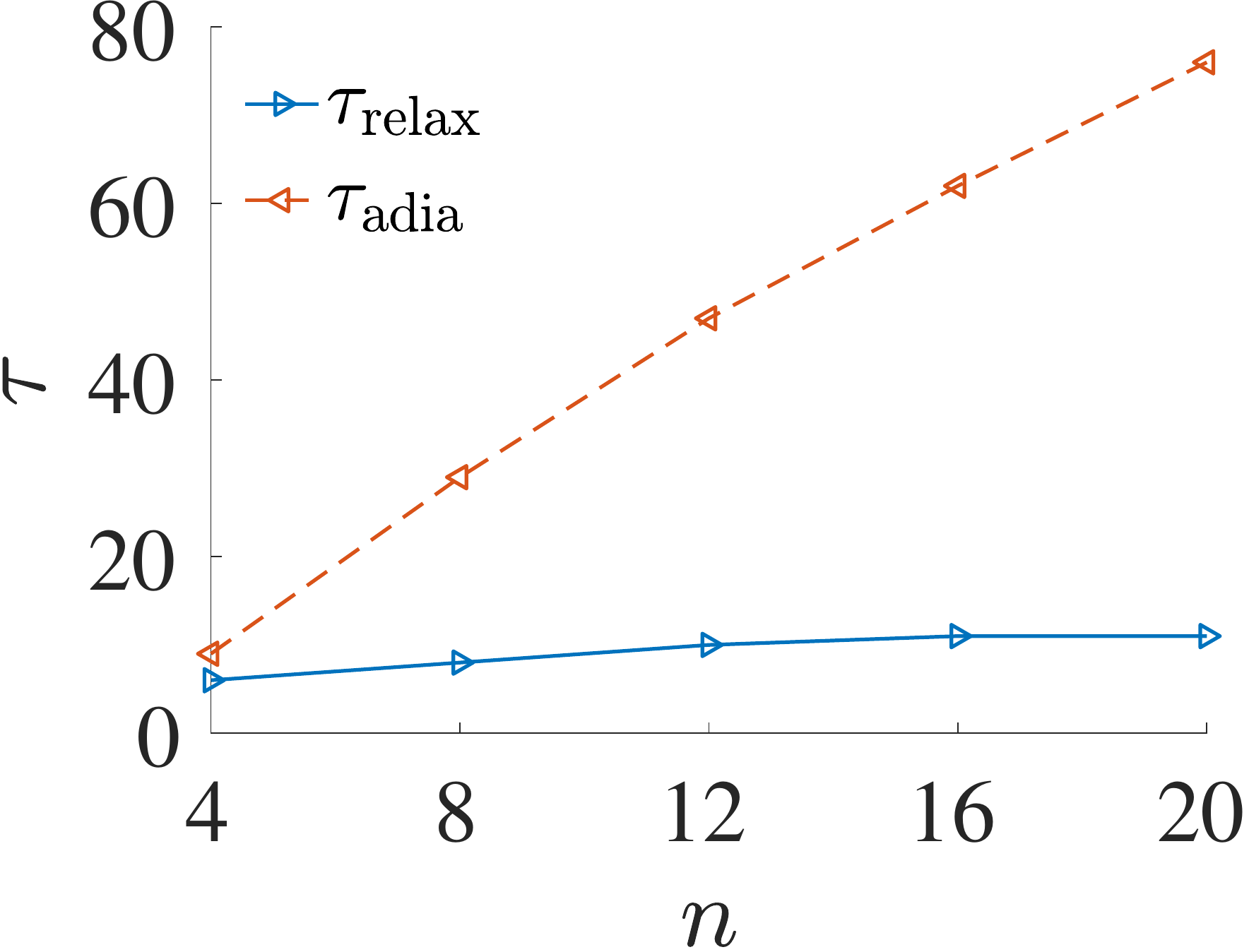}\label{fig:beta01}}
   \caption{(Color online) (a) Scaling of the estimates given by Eqs.~\eqref{eq:B_T0a} (denoted by $B_{\mathrm{int}}$) and \eqref{eq:B_T0b} (denoted by $B_{\max}$), in the low temperature setting ($\beta=10$) and $g^2 = 1$. Also shown is $\tau_{\epsilon}$, the minimum time such that the TND from the instantaneous Gibbs state is $\leq \epsilon= 10^{-2}$ $\forall s\in [0,1]$. This quantity and $B_{\mathrm{int}}$ exhibit very similar scaling. (b) Time to reach a TND of $\epsilon = 10^{-2}$ for adiabatic preparation and relaxation, with $g^2=10^{-2}$ and $\beta = 0.1$. At this relatively high temperature relaxation scales better than adiabatic preparation. Same units as in Fig.~\ref{fig:spike}.}
    \label{fig:11} 
\end{figure}

We have seen that the prediction of the adiabatic preparation TTSS
given by Eq.~\eqref{eq:T_adia} fails for the \qt{spike} problem (in
that it predicts a TTSS increasing with $n$). What
about the $T\to 0$ estimates given by Eqs.~\eqref{eq:B_T0a} and
\eqref{eq:B_T0b}?
We 
next
show that, provided 
appropriate
care is taken in the definition of 
the
TTSS, 
both equations,
and in particular Eq.~\eqref{eq:B_T0a},
provide excellent agreement with the TTSS. 
First, let us recall that
when using
the rescaled variable $s=t/\tau$, we are preparing the
steady state at $s=1$ 
while evolving
at a speed (rate) $1/\tau$. Clearly, the
procedure can be generalized to prepare the state at
$s=s_{\mathrm{fin}}$. A common feature of the bounds in Eqs.~\eqref{eq:B_T0a} and
\eqref{eq:B_T0b} is that they are both monotonically increasing in
$s_{\mathrm{fin}}$ [since $\varepsilon(s)$ is a positive
function]. This means that if we use
Eqs.~\eqref{eq:B_T0a} and \eqref{eq:B_T0b} to estimate $\tau$, then we are
guaranteed that $%
\frac{1}{2}
\Vert \rho(s) - \rho_\mathrm{SS}(s) \Vert%
_1
 \le
\epsilon$ for all $s\le s_\mathrm{fin}$.

In order to enforce a fair comparison,
let us define $\tau_{\epsilon}$ as the minimum time such that the TND from the instantaneous steady state state is always at most $\epsilon$ throughout the entire evolution, i.e.:
\beq
\tau_{\epsilon} = \min_{\tau} \left\{\frac{1}{2}\|\rho_{\mathrm{adia}}(t)-\rho_{\mathrm{SS}}(t)\|_1 \leq \epsilon\quad \forall t\in[0,\tau]\right\}\ .
\eeq

We show the behavior of this new TTSS in
Fig.~\ref{fig:adiabaticSpike}, where it can be seen to agree very well
with the prediction of Eq.~\eqref{eq:B_T0a}. Thus, we may conclude
that the reason that Eq.~\eqref{eq:B_T0a} [and hence also
  Eq.~\eqref{eq:B_T0b}, and in principle also Eq.~\eqref{eq:T_adia}]
does not capture the scaling of $\tau_{\mathrm{adia}}$ for the
\qt{spike} problem is that the latter enforces a small TND only at the
end of the evolution, while the former enforces a small TND throughout
the entire evolution. It is reasonable to conjecture that this
conclusion is valid well beyond the \qt{spike} problem. Moreover, note
that the scaling of $\tau_{\mathrm{adia}}$ is polynomial, so that even
with this stricter notion of adiabatic preparation, relaxation [which
  scales exponentially for the \qt{spike} problem --
  Fig.~\ref{fig:spike1}] is still bested.

\subsubsection{The high temperature case}

To conclude our discussion of the \qt{spike} problem, we show in Fig.~\ref{fig:beta01} that for sufficiently high temperatures, the relaxation process becomes more efficient than adiabatic preparation.  This is unsurprising, since the spike energy barrier is only a hinderance to thermal relaxation if 
crossing it is required in order 
to be $\epsilon$-close to the thermal state.  Therefore, we again find that there is range of parameters where adiabatic preparation can be more efficient than thermal relaxation. We leave open the problem of finding the temperature at which they achieve equal scaling, and what is special about that temperature value.

\section{Conclusions}
\label{sec:7}

Adiabatic quantum computing is an analog algorithm that has generated tremendous recent interest \cite{Albash-Lidar:RMP}. 
Its analog nature
makes a comparison of its efficiency with that of
classical algorithms running on digital machines a subtle problem. In this work we
compared adiabatic quantum preparation of steady states of Lindbladians
 with the relaxation process of the same Lindbladians at the
end of the adiabatic path. In some situations the relaxation process
is described by a classical Markov chain. Hence for these
cases we are able to unambiguously compare quantum and classical
preparation times. Alternatively, this setting can be used to describe
the efficiency of realistic implementations of adiabatic quantum
computing
where the goal is steady state preparation.
Moreover, using known results for the mixing times of
Lindbladian generators and the open-system generalization of the
adiabatic theorem, we are also able to estimate such adiabatic and
relaxation-based preparation times. The result of attempting to use these estimates for a comparison is rather disturbing
for aficionados of computation via adiabatic evolution%
: relaxation seems to always be more efficient than adiabatic preparation. If this were true, it would  doom the nascent field of experimental 
AQC and QA,
which would have to be redirected towards building quantum relaxation devices instead. 

However, while this formal analysis is very general, it only provides a worst-case bound.  .  A deeper investigation reveals that the situation is more subtle than is suggested by the relaxation and adiabatic theorem time estimates. First, we found that by considering the adiabatic bound for thermalizing (Davies) generators, we are able to compute the bound in the \emph{low temperature} regime and estimate its leading behavior. The resulting expression \emph{can} in principle be smaller
than the relaxation time, thus redeeming adiabatic preparation. Second, by studying several models, in particular the \qt{spike} problem, for
which a (limited) quantum speed-up relative to simulated annealing is 
known in the closed-system case \cite{Farhi-spike-problem}, we found that \emph{relaxation-assisted} adiabatic preparation dramatically out-scales final-time relaxation, which scales exponentially with problem size, while the former becomes \emph{faster} as the problem size increases. This conclusion remained qualitatively unchanged even after imposing a stricter notion of instantaneous adiabaticity, which forces the system to always evolve close to the instantaneous steady state, in the sense that adiabatic preparation now scales polynomially with system size.  Therefore, we find that if the system is sufficiently close to the final-time steady state before reaching the end of the evolution, then, as might be expected, the final-time gap does not hinder the adiabatic preparation.

These results are encouraging for the adiabatic preparation of steady states, but it should be remembered that adiabatic quantum computing is traditionally concerned with the preparation of ground states. Moreover, for a different model of quasi-free fermionic chains with an integrable, unstructured (i.e., not thermalizing) Lindbladian we found that relaxation outperforms adiabatic preparation, while for a model of a single qubit with a thermalizing
Lindbladian we found mixed results, with adiabatic preparation beating
relaxation only when both the system-bath coupling and the temperature
are sufficiently small. Thus, our work shows that the conditions under
which adiabatic preparation is superior to relaxation are far from
universal, and more work is needed to discover both general principles
and
specific examples for which adiabatic preparation is the preferred
strategy.

\begin{acknowledgments}
This work was supported under ARO Grant No. W911NF-12-1-0523, ARO MURI Grants No. W911NF-11-1-0268 and No. W911NF-15-1-0582, and NSF Grant No. INSPIRE-1551064.
\end{acknowledgments}

\appendix

\section{Adiabatic error at zero temperature}

\label{app:zeroT}

Recall that, with $Q_{0}(t)=\1-P_{0}(t)$,
\beq
S(t)=\lim_{z\to0}Q_{0}(t)[z-\mathcal{L}(t)]^{-1}Q_{0}(t)
\eeq
is the reduced resolvent of $\mathcal{L}$.
As is customary, in order to avoid a proliferation of factors of $\tau$ we switch to the dimensionless time variable $s=t/\tau$. All functions of time become functions of $s$. With prime denoting differentiation with respect to $s$, the constant $B$ from Eqs.~\eqref{eq:adia-SS} and \eqref{eq:C-adia} can then be written as 
\begin{align}
B &=\left\Vert S(1)\rho_{\mathrm{SS}}'(1)\right\Vert _{1} +\left\Vert S(0)\rho_{\mathrm{SS}}'(0)\right\Vert _{1} \nonumber \\
& +\int_{0}^{1}d\sigma\left\Vert S'\rho_{\mathrm{SS}}'+S\rho_{\mathrm{SS}}''\right\Vert _{1}.
\label{eq:B-adia}
\end{align}

Let us estimate the three contributions in Eq.~\eqref{eq:B-adia} for the case of Davies generators in the zero temperature limit.
Let us assume that the Hamiltonian spectrum is non-degenerate. 
In the zero temperature limit, $\rho_{\mathrm{SS}}\to|0\rangle\langle0|$, where $|0\rangle$
is the Hamiltonian ground state (we drop the time dependence). Since we
are in finite dimension and $H$ is assumed to depend smoothly on $s$,
the limit $T\to 0$ commute with diferentiation with respect to $s$. 

We start by noting that%
\footnote{ 
Differentiate $H\ket{0}=E_0\ket{0}$ to get $(H'\ket{0}-E_0')\ket{0} = (E_0-H)\ket{0'}$. Multiply by $\ketbra{l}$ with $l\neq 0$ to get $\bra{l}H'\ket{0} \ket{l}\bra{0}/(E_0-E_l) = \ketbra{l}\dot{0}\rangle\bra{0}$. Add the h.c.~and sum over all $l\neq 0$ to get $-\sum_{l\neq0}\frac{\langle
    l|H'|0\rangle}{\Delta_{l,0}} |l\rangle\langle
0|+\hc=(|0\rangle\langle0|)'-\ketbra{0}(\langle 0'\vert 0 \rangle +
\langle 0\vert 0' \rangle) = (|0\rangle\langle0|)'$.
}
\begin{equation}
(|0\rangle\langle0|)'=-\sum_{l\neq0}\frac{\langle
    l|H'|0\rangle}{\Delta_{l,0}} |l\rangle\langle 0|+\hc\ 
    \label{eq:rhoprime}  
\end{equation}
Recall that $\mathcal{L}[|n\rangle\langle m|] = \lambda_{n,m} |n\rangle\langle m|$
for $n\neq m$ with $\lambda_{n,m}=-i\Delta_{n,m}- \eta_{n,m}$.
Since $S(|l\rangle\langle 0|)=\lambda_{l,0}^{-1}|l\rangle\langle 0|$,
we see immediately that 
\begin{equation}
\lim_{T\to0}S\rho_{\mathrm{SS}}'=-\sum_{l\neq0}\frac{\langle l|H'|0\rangle}{\Delta_{l,0}\lambda_{l,0}}|l\rangle\langle 0|+\hc\ 
\label{eq:Srhop_easy}
\end{equation}
The matrix in Eq.~\eqref{eq:Srhop_easy} has the form $|0\rangle\langle\phi|+\hc$
where $|\phi\rangle$ is orthogonal to $|0\rangle$. Its trace-norm
is given by $2\left\Vert |\phi\rangle\right\Vert $ so \bes 
\begin{align}
\lim_{T\to0}\left\Vert S\rho_{\mathrm{SS}}'\right\Vert _{1}= & 2\sqrt{\sum_{l>0}\left|\frac{\langle l|H'|0\rangle}{\Delta_{l,0}\lambda_{0,l}}\right|^{2}}\label{eq:norm_unitary}\\
\sim & 2\max_{l\neq0}\left|\frac{\langle l|H'|0\rangle}{\Delta_{l,0}\lambda_{l,0}}\right|\ ,
\end{align}
\ees where in the last step we retained only the leading term. Below we assume that this maximum is attained at $l=1$.  

We now turn our attention to $S\rho''+S'\rho'=\left(S\rho'\right)'$. 
Using Eq.~\eqref{eq:Srhop_easy}, in a few steps one arrives at (from
now on, all equations are intended at zero temperature) 
\begin{align}
S\rho''+S'\rho' & =  \sum_{l\neq0}\frac{\langle l|H'|0\rangle}{\Delta_{l,0}}\frac{(\lambda_{l,0})'}{\lambda_{l,0}^{2}}|l\rangle\langle0|\notag \\
\label{eq:A6}
 & -  \sum_{l\neq0}\frac{\langle l|H'|0\rangle}{\Delta_{l,0}}\frac{1}{\lambda_{l,0}}(|l\rangle\langle0|)'\\
 & -  \sum_{l\neq0}\partial_{s}\left(\frac{\langle l|H'|0\rangle}{\Delta_{l,0}}\right)\frac{1}{\lambda_{l,0}}|l\rangle\langle0|+\hc\notag
\end{align}

The term in the second line is: 
\begin{eqnarray}
(|l\rangle\langle0|)' & = & -\sum_{\stackrel{m\neq l}{m\neq0}}\frac{\langle m|H'|l\rangle}{\Delta_{m,l}}|m\rangle\langle0|-\sum_{\stackrel{m\neq l}{m\neq0}}\frac{\langle0|H'|m\rangle}{\Delta_{m,0}}|l\rangle\langle m|\nonumber \\
 &  & -\frac{\langle0|H'|l\rangle}{\Delta_{0,l}}|0\rangle\langle0|-\frac{\langle0|H'|l\rangle}{\Delta_{l,0}}|l\rangle\langle l|\nonumber \\
 &  & +|l\rangle\langle0|\left(\langle l|l'\rangle+\langle0'|0\rangle\right)\label{eq:l0_prime}
\end{eqnarray}
Using the chain rule repeatedly and noting that $(E_{l})'=\langle l|H'|l\rangle$
we obtain for the third line of Eq.~\eqref{eq:A6}:
\begin{align}
\partial_{s}\left(\frac{\langle l|H'|0\rangle}{\Delta_{l,0}}\right) & =-\sum_{\stackrel{m\neq l}{m\neq0}}\frac{\langle l|H'|m\rangle}{\Delta_{m,l}}\frac{\langle m|H'|0\rangle}{\Delta_{l,0}}\nonumber \\
 & -\sum_{\stackrel{m\neq l}{m\neq0}}\frac{\langle l|H'|m\rangle}{\Delta_{l,0}}\frac{\langle m|H'|0\rangle}{\Delta_{l,0}}\nonumber \\
 & +\frac{\langle l|H''|0\rangle}{\Delta_{l,0}}-2\frac{\langle l|H'|0\rangle}{\left(\Delta_{l,0}\right)^{2}}\left(\langle l|H'|l\rangle-\langle0|H'|0\rangle\right)\nonumber \\
 & +\frac{\langle l|H'|0\rangle}{\Delta_{l,0}}\left(\langle
l'|l\rangle+\langle0|0'\rangle\right). \label{eq:partial_s} 
\end{align}

The Berry's connection terms (of the form 
$\langle l' | l \rangle$
or complex conjugate) in Eqs.~\eqref{eq:l0_prime}
and \eqref{eq:partial_s} cancel out exactly and we arrive at:
\begin{align}
(S\rho')' & =\sum_{l\neq0}\frac{\langle l|H'|0\rangle}{\Delta_{l,0}}\frac{(\lambda_{l,0})'}{\lambda_{l,0}^{2}}|l\rangle\langle0|\notag\\
 & +\sum_{l\neq0}\sum_{{m\neq 0,l}}\frac{\langle l|H'|0\rangle}{\Delta_{l,0}}\frac{\langle m|H'|l\rangle}{\Delta_{m,l}}\frac{1}{\lambda_{l,0}}|m\rangle\langle0|\notag\\
 & +\sum_{l\neq0}\sum_{{m\neq 0,l}}\frac{\langle l|H'|0\rangle}{\Delta_{l,0}}\frac{\langle0|H'|m\rangle}{\Delta_{m,0}}\frac{1}{\lambda_{l,0}}|l\rangle\langle m|\notag\\
 & +\sum_{l\neq0}\left|\frac{\langle l|H'|0\rangle}{\Delta_{l,0}}\right|^{2}\frac{1}{\lambda_{l,0}}|l\rangle\langle l|\notag\\
 & -\sum_{l\neq0}\left|\frac{\langle l|H'|0\rangle}{\Delta_{l,0}}\right|^{2}\frac{1}{\lambda_{l,0}}|0\rangle\langle0|\notag\\
 & +\sum_{l\neq0}\sum_{{m\neq 0,l}}\frac{\langle l|H'|m\rangle}{\Delta_{m,l}}\frac{\langle m|H'|0\rangle}{\Delta_{l,0}}\frac{1}{\lambda_{l,0}}|l\rangle\langle0|\notag\\
 & +\sum_{l\neq0}\sum_{{m\neq 0,l}}\frac{\langle l|H'|m\rangle}{\Delta_{l,0}}\frac{\langle m|H'|0\rangle}{\Delta_{m,0}}\frac{1}{\lambda_{l,0}}|l\rangle\langle0|\notag\\
 & -\sum_{l\neq0}\frac{\langle l|H''|0\rangle}{\Delta_{l,0}}\frac{1}{\lambda_{l,0}}|l\rangle\langle0|\notag\\
 & +2\sum_{l\neq0}\frac{\langle
  l|H'|0\rangle \left(\langle
  l|H'|l\rangle-\langle0|H'|0\rangle\right)}{\Delta_{l,0}^{2}
\lambda_{l,0}} |l\rangle\langle0|+\hc\notag\\
\end{align}
The third and the fourth terms can be combined and, after some other
minor adjustments, we finally arrive at: 
\begin{align}
(S\rho')' & =-\sum_{l\neq0}\left|\frac{\langle
  l|H'|0\rangle}{\Delta_{l,0}}\right|^{2}\frac{1}{\lambda_{l,0}}|0\rangle\langle0|\nonumber \\
& +\sum_{l\neq0}\sum_{m\neq0}\frac{\langle l|H'|0\rangle}{\Delta_{l,0}}\frac{\langle0|H'|m\rangle}{\Delta_{m,0}}\frac{1}{\lambda_{l,0}}|l\rangle\langle m|\nonumber \\
 & +\sum_{l\neq0}\sum_{{m\neq 0,l}}\frac{\langle
  m|H'|0\rangle\langle l|H'|m\rangle}{\Delta_{l,m}} \times\nonumber \\
& \times \left(\frac{1}{\Delta_{m,0}\lambda_{m,0}}-\frac{1}{\Delta_{l,0}\lambda_{l,0}}\right)|l\rangle\langle0|\nonumber \\
 & +\sum_{l\neq0}\frac{\langle l|H'|0\rangle}{\Delta_{l,0}}\frac{(\lambda_{l,0})'}{\lambda_{l,0}^{2}}|l\rangle\langle0|-\sum_{l\neq0}\frac{\langle l|H''|0\rangle}{\Delta_{l,0}}\frac{1}{\lambda_{l,0}}|l\rangle\langle0|\nonumber \\
 & +\sum_{l\neq0}\sum_{{m\neq 0,l}}\frac{\langle l|H'|m\rangle}{\Delta_{l,0}}\frac{\langle m|H'|0\rangle}{\Delta_{m,0}}\frac{1}{\lambda_{l,0}}|l\rangle\langle0|\nonumber \\
 & +2\sum_{l\neq0}\frac{\langle
  l|H'|0\rangle \left(\langle
  l|H'|l\rangle-\langle0|H'|0\rangle\right)}{\Delta_{l,0}^{2}
\lambda_{l,0}} |l\rangle\langle0|+\hc
\label{eq:error_total}
\end{align}
The third term in Eq.~\eqref{eq:error_total} appears problematic, i.e., as if it may diverge as an inverse gap (in fact, a gap in the middle of the spectrum) when $E_{l}\approx E_{m}$. However in this case also the term in parenthesis vanishes.
Denoting $\Delta_{m}:=\Delta_{m,0}$, one has, formally
(differentaiting with respect to $m$):
\beq
``{\lim_{m\to l}} (\ast) "=\langle l|H'|0\rangle\langle
l|H'|l\rangle\frac{\lambda_{l,0}\partial_m
  \Delta_{l}+\Delta_{l}\partial_m \lambda_{l,0}}{\Delta_{l}^{2}\lambda_{l,0}^{2}\partial_m\Delta_{l}}\ ,
\eeq
where $(\ast)$ denotes the coefficient of $|l\rangle\langle0|$ in the third and fourth line of Eq.~\eqref{eq:error_total}. Hence this term is bounded when $m\to l$.
Instead, when $m\to0$ (or $l\to0$) this term {[}as well as other
terms in Eq.~\eqref{eq:error_total}{]} does diverge if not compensated
by a vanishing of the matrix elements of $H'$. We conclude that the
largest contribution to this term comes from terms with $\Delta_m\approx0$
(or $\Delta_l\approx0$). 

Note that only particular Lindbladian gaps
$\lambda_{l,0}$ enter the above expression. This is not a priori
obvious. Indeed, if one computes the contributions $S'\rho'$, $S\rho''$
separately, using 
$S'=S^{2}\mathcal{L}'P+P\mathcal{L}'S^{2}-S\mathcal{L}'S$, 
as shown in \cite{venuti_adiabaticity_2016}, one obtains contribution
also from other Lindbladian eigenvalues such as $\lambda_{n,m}$ and
$\lambda_m$. It turns out, however, that such contributions exactly cancel
out once summed. 

Considering Eq.~\eqref{eq:error_total}, it is evident that the
operator $S\rho''+S'\rho'$ has the following form
\begin{equation}
S\rho''+S'\rho'=A|0\rangle\langle0|+|\xi\rangle\langle\eta|+|\eta\rangle\langle\xi|+|0\rangle\langle\phi|+|\phi\rangle\langle0|
\end{equation}
With
\bes
\begin{align}
A & =  -2\sum_{l\neq0}\left|\frac{\langle l|H'|0\rangle}{\Delta_{l,0}}\right|^{2}\mathrm{Re}\left(\frac{1}{\lambda_{l,0}}\right)\\
|\xi\rangle & =  \sum_{l\neq0}\frac{\langle l|H'|0\rangle}{\Delta_{l,0}}\frac{1}{\lambda_{l,0}}|l\rangle\\
|\eta\rangle & =  \sum_{m\neq0}\frac{\langle m|H'|0\rangle}{\Delta_{m,0}}|m\rangle,
\end{align}
\ees
and $|\phi\rangle$ can be read off from the last four lines of Eq.~\eqref{eq:error_total}.
Note that $|\xi\rangle,|\eta\rangle,|\phi\rangle$ are all orthogonal
to $|0\rangle$. Therefore, at zero temperature $S\rho''+S'\rho'$ is just
a rank-four matrix. Unfortunately, although in principle it is possible
to compute its eigenvalues, these are roots of a fourth order polynomial.
Moreover we are interested in its leading contribution. We \emph{assume} that the leading term in the above expressions comes
from terms with $\Delta_{1,0}^{-2}$ and discard all the other terms. In this approximation

\begin{equation}
(S\rho')' \approx A|0\rangle\langle0|+ b |0\rangle\langle 1|+\overline{b} |1\rangle\langle0| + c |1\rangle\langle1|
\end{equation}

with 
\begin{align}
b &= 2\frac{\langle1|H'|0\rangle}{\Delta_{1,0}^{2}}
\frac{\langle1|H'|1\rangle-\langle0|H'|0\rangle}{\lambda_{1,0}} \\
c &= 2\left|\frac{\langle1|H'|0\rangle}{\Delta_{1,0}}\right|^{2}\left|\mathrm{Re}\left(\frac{1}{\lambda_{1,0}}\right)\right|
\end{align}

Taking the trace-norm we obtain:
\begin{align} 
\label{eqt:ConstantBnoIntegral}
\left\Vert S\rho''+S'\rho'\right\Vert _{1} &\approx \sqrt{ (A-c)^2 + 4 \vert b \vert^2}  \le \vert A-c \vert + 2 \vert b \vert \nonumber \\
& = 4\frac{\left|\langle1|H'|0\rangle\right|}{\Delta_{1,0}^{2}}\left|\frac{\langle1|H'|1\rangle-\langle0|H'|0\rangle}{\lambda_{1,0}}\right| \nonumber \\
& \qquad +4\left|\frac{\langle1|H'|0\rangle}{\Delta_{1,0}}\right|^{2}\left|\mathrm{Re}\left(\frac{1}{\lambda_{1,0}}\right)\right|.
\end{align}
The above equation is manifestly continuous in the adiabatic time
variable $s$, because
eigenvalues are continuous and $H$ is smooth. Using
Eq.~\eqref{eq:B-adia} and  $\int_0^1 |f(\sigma)| d\sigma \le
\max_{\sigma\in[0,1]} |f(\sigma)|$ (valid for continuous $f$), one obtains
Eqs.~\eqref{eq:eps(t)} and \eqref{eq:B_T0}, after switching back to the
time variable $t$. 

\section{Finite temperature corrections}
\label{app:finiteT}

Let us comment on the corrections to the above results due to a small, positive temperature. The leading corrections to $\rho_\mathrm{SS}$ have the form $Q \exp{(-\Delta_{1,0}/T})$ where $Q$ is temperature independent. This means that corrections to the derivatives $\rho_\mathrm{SS}^{(n)}$ are $O\left[T^{-n} \exp{(-\Delta_{1,0}/T)}\right]$. In other words Eq.~\eqref{eqt:ConstantBnoIntegral} is correct up to $O\left[T^{-2} \exp{(-\Delta_{1,0}/T)}\right]$. When plugging this result into Eq.~\eqref{eq:B-adia} it is certainly possibly to bound the error as
$O\left[T^{-2} \exp{(-\Delta_{1,0}^{\mathrm{min}}/T)}\right]$ where $\Delta_{1,0}^{\mathrm{min}} = \min_{t\in[0,\tau]} \Delta_{1,0}(t)$.

%

\end{document}